\documentclass[onefignum,onetabnum]{siamonline220329}

\usepackage{lipsum}
\usepackage{amsfonts}
\usepackage{amssymb}
\usepackage{graphicx}
\usepackage{epstopdf}
\usepackage{algorithmic}
\ifpdf
  \DeclareGraphicsExtensions{.eps,.pdf,.png,.jpg}
\else
  \DeclareGraphicsExtensions{.eps}
\fi

\usepackage{subcaption}
\renewcommand{\vec}[1]{\mathbf{#1}} 
\renewcommand{\Re}{\mbox{\textit{Re}}} 
\newcommand{\Rep}{\Re_{p}} 

\usepackage{enumitem}
\setlist[enumerate]{leftmargin=.5in}
\setlist[itemize]{leftmargin=.5in}

\newsiamremark{remark}{Remark}
\newsiamremark{hypothesis}{Hypothesis}
\crefname{hypothesis}{Hypothesis}{Hypotheses}
\newsiamthm{claim}{Claim}

\headers{Inertial migration in curved trapezoidal ducts}{B. Harding, Y. Stokes, R. Valani}

\title{Inertial focusing dynamics of spherical particles in curved microfluidic ducts with a trapezoidal cross-section.
\thanks{Submitted to the editors DATE.
\funding{This work was supported by the Australian Research Council (DP200100834) and a Victoria University of Wellington Faculty Research Establishment Grant.}
}}

\author{Brendan Harding\thanks{School of Mathematics and Statistics, Victoria University of Wellington, Wellington, NZ 
  (\email{brendan.harding@vuw.ac.nz}).}
\and Yvonne M. Stokes\thanks{School of Mathematical Sciences, University of Adelaide, Adelaide, South Australia, Australia (\email{yvonne.stokes@adelaide.edu.au}, \email{rahil.valani@adelaide.edu.au})}
\and Rahil N. Valani\footnotemark[3]
}

\usepackage{amsopn}

\ifpdf
\hypersetup{
  pdftitle={Inertial focusing in curved trapezoidal ducts},
  pdfauthor={B. Harding}
}
\fi

\begin{document}

\maketitle

\begin{abstract}
Inertial focusing in curved microfluidic ducts exploits the interaction of drag force from the Dean flow with the inertial lift force to separate particles or cells laterally across the cross-section width according to their size.
Experimental work has identified that using a trapezoidal cross-section, as opposed to a rectangular one, can enhance the sized based separation of particles/cells over a wide range of flow rates.
Using our model, derived by carefully examining the way the Dean drag and inertial lift forces interact at low flow rates, we calculate the leading order approximation of these forces for a range of trapezoidal ducts, both vertically symmetric and non-symmetric, with increasing amount of skew towards the outside wall.
We then conduct a systematic study to examine the bifurcations in the particle equilbira that occur with respect to a shape parameter characterising the trapezoidal cross-section. 
We reveal how the dynamics associated with particle migration are modified by the degree of skew in the cross-section shape, and show the existence of cusp bifurcations (with the bend radius as a second parameter).
Additionally, our investigation suggests an optimal amount of skew for the trapezoidal cross-section for the purposes of maximising particle separation over a wide range of bend radii.
\end{abstract}




\section{Introduction}

Inertial lift is a phenomenon that causes finite sized particles suspended in a fluid flow to migrate across streamlines.
Its experimental identification by Segr\'e and Silberberg \cite{SegreSilberberg1961} sparked a significant number of modelling efforts spanning several decades 
\cite{Asmolov1999,HoLeal1974,Hogg1994,Saffman1965,SchonbergHinch1989}. 
The cross-stream migration is generated through the disturbance of the fluid by the finite size of the particle.
Generally speaking, this is a difficult problem to model and as such most studies consider a relatively simple setup, the most common being a single spherical particle suspended in a simple shear or Poisseuille flow bounded by two plane parallel walls (i.e. confined in only one spatial direction).
In contrast, the applications in which inertial lift is exploited use enclosed geometries and it is only relatively recently that models of such setups have been explored 
\cite{HardingStokesBertozzi2019,HoodLeeRoper2015,MatasMG2009}. 
The migration of particles or cells in the plane of a cross-section driven, at least in part, by inertial lift is herein referred to as inertial migration.

Inertial migration of particles and cells is of great interest to the microfluidics community and has led to a resurgence of interest in the phenomenon.
There are numerous applications related to separation of particles having different physical characteristics, such as size and/or density.
Examples include the isolation of circulating tumor cells from a blood sample \cite{WarkianiEtal2016}, the filtering of bacteria and/or pollutants from a water sample \cite{TothEtal2016}, and the extraction of metals and minerals \cite{PriestEtal2011,YinEtal2013}.
Designing optimal microfluidic devices for a diverse range of applications requires further advances in modelling to aid design.
This is particularly the case in microfluidic devices featuring a curved duct geometry because the presence of a secondary flow in the cross-sectional plane significantly complicates the migration dynamics \cite{GossettDiCarlo2009}.

To study this problem, the fluid flow is typically modelled by the Navier--Stokes equations explicitly coupled to the particle motion through boundary conditions, while the particle motion is driven by the hydrodynamic force and torque (and gravitational force where appropriate).
By using sophisticated numerical methods and sufficiently powerful computing resources it is possible to directly simulate the motion of one or more particles suspended in fluid flow.
However, direct simulation of complex devices remains burdensome and ultimately provides limited insight into the global dynamics associated with inertial particle migration \cite{LiuEtal2016}.
The dynamics can be studied in greater detail by utilising a quasi-static approximation of the same setup and computing the forces driving particle motion via a regular perturbation expansion 
\cite{HardingStokesBertozzi2019,HoodLeeRoper2015}. 
The resulting particle migration model, expressible as ordinary differential equations for the coordinates of the centre of the particle, facilitates an in-depth study of the many parameters that can influence the migration dynamics.
Key parameters that have been explored to date include particle size, the aspect ratio of rectangular cross-sections, and the bend radius of the duct (in the case of curved duct geometries) \cite{valani2021_b}. 

An interesting line of enquiry is how the cross-section shape influences the dynamics of inertial migration more generally.
This has been partly driven by an interest in trapezoidal cross-sections which have been shown to enhance the efficiency of size based particle/cell separation in comparison to rectangular cross-sections
\cite{AkbarnatajEtal2023,LeeEtal2015,RafeieEtal2019,WarkianiEtal2014}. 
There has been an attempt to estimate the location of stable equilibria when stitching together different cross-sectional shapes \cite{RafeieEtal2019_2}, but this is largely based on heuristics and, while there is some qualitative agreement with experimental work, it is unclear how well it would agree with more detailed quantitative studies.
A more recent paper developed a numerical approach inspired by regularised Stokeslets and does a good job at estimating stable focusing locations of sufficiently small particles suspended in flow through straight duct geometries having arbitrary cross-sectional shape \cite{ChristensenEtal2022}.
It is currently unclear if this approach could be adapted into an efficient methodology for exploring dynamics within curved duct geometries.
There is clearly much demand to understand the influence of cross-sectional shape on inertial migration but, simultaneously, a lack of detailed quantitative studies in the literature.

In this paper we present an in-depth quantitative study of migration dynamics within curved duct geometries having trapezoidal cross-sections, building on the single trapezoidal cross-section considered in \cite{HardingStokesBertozzi2019}.
Our previous work on inertial migration within curved duct geometries having square/rectangular cross-sections demonstrates our model is conducive to a thorough dynamical systems study of the equilibria within the hydrodynamic force field within a cross-sectional plane 
\cite{HaHardingBertozziStokes2022,valani2021_b,ValaniHS2023}. 
These studies have revealed a rich landscape of equilibria configurations and bifurcations with respect to key system parameters that separate different regimes.
By applying the same model and methodology to trapezoidal cross-sections we aim to answer several key questions: 
a) in what ways do particle migration dynamics in curved ducts with trapezoidal cross-sections differ from those in rectangular cross-sections, 
b) are there any interesting bifurcations that occur as the skew in the trapezoidal shape is increased, 
and c) what are the key differences in particle focusing behaviour between trapezoidal cross-sections having a vertically symmetric shape versus those with a `flat bottom'?

Previous studies identified a dimensionless parameter $\kappa$, defined via three length scales (see section \ref{sec:background}), as being fundamental in approximately characterising focusing behaviour at low flow rates in curved ducts with a rectangular cross-section \cite{HardingStokesBertozzi2019}.
In particular, there is an intermediate range of $\kappa$ for which particles migrate to a stable equilibria pair located nearer to the inside wall, but either side of which the stable equilibria pair shifts horizontally towards the centre of the cross-section. 
This ultimately limits the region in which separation of different sized particles occurs to just half of the width of the cross-section nearest the inside wall.
An experimental study which explored a variety of trapezoidal shapes illustrated that particles can be focused nearer to the outside wall by increasing the flow rate and/or using trapezoidal cross-sections which are taller at the outside wall~\cite{RafeieEtal2019}.
Since our interest is in understanding sized based separation at low flow rates, we focus the attention of this paper on studying trapezoidal cross-sections which are taller at the outside wall.

Our results aim to quantify, for a given bend radius, the extent that the trapezoidal cross-section shape, with taller outside wall, pushes features of the flow towards the outer wall and ultimately allows the horizontal location of stable equilibria, for a range of particle sizes, to cover a greater range of the cross-sectional width.
We also find there is an optimal amount of skew in the trapezoidal shape after which different sized particles become less well separated.
Moreover, there is a cusp bifurcation that occurs with respect to the trapezoidal shape and curvature parameters, which results in two well separated stable equilibria pairs over a certain range of particle sizes and bend radii.
We also find that the main difference between the vertically symmetric and flat-bottomed trapezoidal shapes is that the latter produces a small offset in the location of each equilibrium in each stable pair.
This offset increases with the trapezoidal shape parameter and may ultimately become counter-productive in practical applications.
Apart from this offset, the general dynamical behaviour otherwise appears to be qualitatively similar, which suggests a degree of robustness in the dynamics.

The paper is organised as follows. 
We begin in section \ref{sec:background} with a brief description of our inertial migration model and describe how it has been adapted for the case of trapezoidal cross-sections.
Following this, in section \ref{sec:strap}, we explore the dynamical landscape of particle migration in the case of vertically symmetric trapezoidal cross-sections.
Then, in section \ref{sec:utrap} we examine the case of flat-bottomed trapezoidal cross-sections with a focus on how this compares to the vertically symmetric case.
Lastly, we end with some conclusions and comments on future directions of this work.

\section{Background}\label{sec:background}

Our methodology for modelling the migration of neutrally buoyant spherical particles at low flow rates, using approximations at leading order in the particle Reynolds number of the inertial lift force and secondary flow drag, has been well established \cite{HardingStokesBertozzi2019}.
This was subsequently applied to study the dynamics of non-neutrally buoyant particles \cite{HardingStokes2020} and then extended to model scenarios with a moderately sized Dean number \cite{HardingStokes2023}.
In order to keep this paper tightly focused we will utilise the original low flow rate model and only consider neutrally buoyant particles.
We provide only a brief outline of the model here, summarising the key assumptions that go into its development and how it is applied in this study.

\begin{figure} 
\centering
\includegraphics{{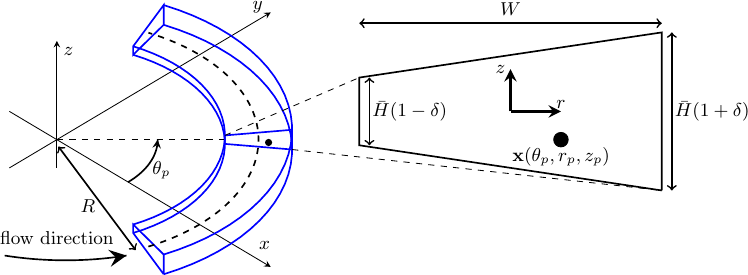}}
\caption{
Portion of a curved duct having a vertically symmetric trapezoidal cross-section and containing a spherical particle located at $\vec{x}_{p}=\vec{x}(\theta_{p},r_{p},z_{p})$. 
The enlarged view of the cross-section containing the particle illustrates the origin of the local $r,z$ coordinates at the centre of the duct.
The bend radius $R$ is measured to the centre-line of the duct and is quite small here for illustration purposes.
We neglect the transition regions near the inlet/outlet of the duct.
Adapted from \cite{HardingStokes2020}.}\label{fig:duct_setup}
\end{figure}

The general setup is depicted in figure~\ref{fig:duct_setup} and is similar to that in previous studies \cite{HardingStokesBertozzi2019} apart from the specific cross-section shape.
Coordinates within the curved duct are most readily described using a cylindrical coordinate system, specifically
\begin{equation}\label{eqn:cc}
\mathbf{x}(r,\theta,z) = (R+r)\cos(\theta)\mathbf{i}+(R+r)\sin(\theta)\mathbf{j}+z\mathbf{k} \,,
\end{equation}
where $R$ is the bend radius of the duct measured from the origin to the centre of the cross-section, described generically by $(r,z)\in\mathcal{C}$ where $\mathcal{C}$ denotes the two-dimensional cross-section.
Consequently, the duct interior is described by $\mathcal{D}=\{\mathbf{x}(\theta,r,z) \mid (r,z)\in\mathcal{C}\}$.

The trapezoidal cross-sections considered herein may be characterised as having vertical side walls separated by the width $W$ and having height $(1-\delta)\bar{H}$ at the inside wall and $(1+\delta)\bar{H}$ at the outside wall, with $\bar{H}$ denoting the mean height (equivalently, the height at the centre). 
The top and bottom walls which connect the side walls are straight lines and the vertical distance between them at any given 
$W/2\leq r\leq W/2$ is given by 
\begin{equation}\label{eqn:height}
H(r)=(1+2\delta r/W)\bar{H} \,.
\end{equation}
The relative alignment of the inside and outside side walls leads to a number of different possible variations.
Two specific variations will be considered, one in which the centres of the side walls are aligned to produce a vertically symmetric cross-section, and another in which the bottoms of the side walls are aligned producing an asymmetric `flat-bottomed' cross-section.
The latter asymmetric family of cross-sections resembles the shape considered in a number of experimental studies \cite{RafeieEtal2019,WarkianiEtal2014}.
This flat-bottomed shape is of importance as it is more easily manufactured via existing techniques (although by no means simple).
 
The vertically symmetric and flat-bottomed variations will be denoted by $\mathcal{C}_s$ and $\mathcal{C}_a$, respectively, and may be formally described as
\begin{subequations}\label{eqn:trap_cs}
\begin{align}
\mathcal{C}_s &:= \left\{(r,z) \,\middle|\, -\frac{W}{2}\leq r\leq\frac{W}{2} \,,\quad  -\frac{H}{2}\left(1+\delta\frac{2r}{W}\right)\leq z\leq\frac{H}{2}\left(1+\delta\frac{2r}{W}\right)\right\} \,, \label{eqn:strap_cs}\\
\mathcal{C}_a &:= \left\{(r,z) \,\middle|\, -\frac{W}{2}\leq r\leq\frac{W}{2} \,,\quad  -\frac{H}{2}\leq z\leq\frac{H}{2}\left(1+2\delta\frac{2r}{W}\right)\right\} \,. \label{eqn:utrap_cs}
\end{align}
\end{subequations}
The effect of the shape parameter $\delta$ on both variations is briefly summarised as:
\begin{itemize}
\item $\delta=0$ describes a rectangular cross-section (in which case $\mathcal{C}_s=\mathcal{C}_a$),
\item $0<\delta<1$ describes a trapezoidal cross-section with outside wall (relative to the bend) taller than the inside wall,
\item $-1<\delta<0$ describes a trapezoidal cross-section with outside wall shorter than the inside wall,
\item $\delta=\pm1$ describes a triangular cross-section (with inside wall having zero height when $\delta=+1$ and the outside wall having zero height when $\delta=-1$).
\end{itemize}
In this paper we are primarily interested in trapezoidal cross sections having $0\leq \delta\leq0.4$ and with width $W$ larger than the average/central height $\bar{H}$.
The setup for a vertically symmetric cross-section is depicted in Figure~\ref{fig:duct_setup}. 

Aside from cross-sectional shape, the model of particle migration used herein is identical to that in previous work in which a single trapezoidal cross-section was considered \cite{HardingStokesBertozzi2019}.
Briefly, the particle is assumed to be held at a fixed location in the cross-section, having an axial velocity and spin such that it is at equilibrium with the surrounding fluid flow, excepting for a hydrodynamic force on it directed within the cross-sectional plane.
The force is computed and the particle migration velocity is then inferred such that the drag on the particle due to its motion is equal and opposite to it.
The hydrodynamic forcing effectively consists of two parts, the first is the inertial lift force induced by the disturbance to the fluid motion caused by the finite size of the particle, and the second is the drag force induced by the Dean flow vortices which develop in curved duct flow \cite{Dean1927}.
Leading order contributions of these two forces are obtained via the use of regular perturbation expansions with respect to particle Reynolds number $\Rep$ and Dean number $K$, respectively, each defined in \eqref{eqn:params} below.
The migration model ultimately consists of a system of first order ordinary differential equations.

Let $a$ denote the particle radius, 
$r_p,z_p$ denote the centre of the particle within the cross-section,
$\ell=\min\{W,\bar{H}\}$ denote the duct length scale (always $\ell=\bar{H}$ in this paper), 
$U_m$ denote the maximum axial flow velocity, 
$R$ denote the bend radius of the duct, 
$\rho$ denote the (uniform) fluid density, and
$\mu$ denote the (uniform) fluid viscosity.
The particle is assumed to be neutrally buoyant, i.e. having density equal to that of the fluid.
We introduce the dimensionless parameters
\begin{align}\label{eqn:params}
\epsilon=\frac{\ell}{2R} \,,\quad 
\alpha=\frac{2a}{\ell} \,,\quad 
\kappa=\frac{4\epsilon}{\alpha^3} \,,\quad 
\Re=\frac{\rho}{\mu}U_m\frac{\ell}{2} \,,\quad 
\Rep=\frac{1}{2}\Re\alpha^2 \,,\quad 
K=\epsilon\Re^2 \,. 
\end{align}
The migration model, neglecting the axial motion of the particle, can ultimately be expressed as a balance between Stokes drag and the sum of the inertial lift and secondary drag. 
Specifically, in dimensionless form, the equations are
\begin{subequations}\label{eqn:ode_model}
\begin{align}
\hat{D}_r(\hat{r}_p,\hat{z}_p)\frac{d \hat{r}_p}{d\hat{t}} &= \hat{L}_r(\hat{r}_p,\hat{z}_p)+\kappa\hat{S}_r(\hat{r}_p,\hat{z}_p)
&\Longrightarrow&&
\frac{d \hat{r}_p}{d\hat{t}} &= \frac{\hat{L}_r(\hat{r}_p,\hat{z}_p)+\kappa\hat{S}_r(\hat{r}_p,\hat{z}_p)}{\hat{D}_r(\hat{r}_p,\hat{z}_p)}\,, \\
\hat{D}_z(\hat{r}_p,\hat{z}_p)\frac{d \hat{z}_p}{d\hat{t}} &= \hat{L}_z(\hat{r}_p,\hat{z}_p)+\kappa\hat{S}_z(\hat{r}_p,\hat{z}_p)
&\Longrightarrow&&
\frac{d \hat{z}_p}{d\hat{t}} &= \frac{\hat{L}_z(\hat{r}_p,\hat{z}_p)+\kappa\hat{S}_z(\hat{r}_p,\hat{z}_p)}{\hat{D}_z(\hat{r}_p,\hat{z}_p)} \,,
\end{align}
\end{subequations}
where, for $\ast=r$ or $z$, 
$\hat{L}_\ast$ is the inertial lift contribution, $\hat{S}_\ast$ is the Dean flow drag contribution
and $\hat{D}_\ast$ is the drag coefficient.
The dimensionless parameter $\kappa$ (defined in \eqref{eqn:params} above) expresses the relative scaling of the Dean flow drag to the inertial lift.
Each of the terms in \eqref{eqn:ode_model} has been non-dimensionalised according to:
\begin{align*}
r_p &= a\hat{r}_p \,, & z_p &= a\hat{z}_{p} \,, & t &= \frac{1}{\Rep}\frac{\ell}{U_m}\hat{t} \,, \\
L_\ast &= \rho U_m^2\frac{a^4}{\ell^2}\hat{L}_\ast \,, & S_\ast &= \epsilon \Re U_m\hat{S}_\ast \,, & D_\ast &= \mu a\hat{D}_\ast \,. 
\end{align*}

The fields $\hat{L}_\ast,\hat{S}_\ast,\hat{D}_\ast$ have an implicit dependence on the trapezoidal shape parameter $\delta$ in addition to the dimensionless parameters $\alpha$, $\epsilon$ and $\Re_p$.
Examples of these fields will be provided in section~\ref{sec:strap}.
The dependence on $\epsilon$ is weak, particularly for $\epsilon<0.1$.
The dependence on $\alpha$ is not particularly strong but has some important implications for applications involving size based separation.
Our asymptotic model neglects the effects of $\Re_p$, apart from those explicitly accounted for in the scaling.
This limits the applicability of our model to low flow rate scenarios, but our experience suggests this still provides significant insight into the more general problem.

The $\hat{L}_\ast$, $\hat{S}_{\ast}$ and $\hat{D}_{\ast}$ fields are carefully estimated from numerical solutions of a problem derived from the Navier--Stokes equations.
Specifically, the fluid motion is governed by
\begin{align*}
\nabla\cdot\left(-p\mathbb{I}+\mu(\nabla\mathbf{u}+\nabla\mathbf{u}^T)\right) &=\rho\left(\frac{\partial \mathbf{u}}{\partial t}+\mathbf{u}\cdot\nabla\mathbf{u}\right) \,, &&\mathbf{x}\in\mathcal{F} \,, \\
\nabla\cdot\mathbf{u} &=0 \,, &&\mathbf{x}\in\mathcal{F} \,, \\
\mathbf{u} &=\mathbf{0} \,, &&\mathbf{x}\in\partial\mathcal{D} \,, \\
\mathbf{u} &=\mathbf{u}_p+\boldsymbol{\Omega}_p{\times}(\mathbf{x}-\mathbf{x}_p) \,, &&\mathbf{x}\in\partial\mathcal{P} \,.
\end{align*}
Here $\mathcal{F}$ denotes the fluid domain, $\partial\mathcal{D}$ denotes the duct boundary, $\partial\mathcal{P}$ denotes the particle boundary, $\mathbf{u}$ is the fluid velocity, $p$ is the pressure, $\mathbf{u}_p$ denotes the particle velocity and $\boldsymbol{\Omega}_p$ denotes its spin (angular velocity).
It is assumed that (neutrally buoyant) particle motion is driven solely by the hydrodynamic force and torque exerted by the fluid, for example
\begin{align*}
\mathbf{F}_p = \int_{\Gamma_p}-\mathbf{n}\cdot\left(-p\mathbb{I}+\mu(\nabla\mathbf{u}+\nabla\mathbf{u}^T)\right)\,dS
\end{align*}
describes the force with $\Gamma_p$ being the particle surface and $\mathbf{n}$ the outward directed unit normal vector.
The derivation of our model involves moving to a rotating reference frame in which the flow is approximately steady and a quasi-static approximation is applicable, decomposing the pressure and velocity fields into background flow and disturbance flow components, and applying a regular perturbation expansion to each component under the assumption that both the Dean number and particle Reynolds number are suitably small, respectively.
For a discussion of the validity of these assumptions and the specifics of the estimation of each component we refer the reader to the original work \cite{HardingStokesBertozzi2019}.

Estimation of the $\hat{L}_\ast$, $\hat{S}_{\ast}$ and $\hat{D}_{\ast}$ fields requires numerical computations at sampling locations $(r_p,z_p)$. 
At each of these we first balance the force on the particle directed down the main axis and the torque components orthogonal to the main axis.
In doing so we determine the corresponding axial and angular velocity components of the particle. 
Only after these have been determined can the appropriate values of $\hat{L}_\ast$, $\hat{S}_{\ast}$ and $\hat{D}_{\ast}$ be estimated at a given $(r_p,z_p)$.
However, once these fields have been estimated, we can ultimately ignore the specific values of the axial and angular velocities if we are only interested in examining the cross-sectional dynamics.
As the cross-sectional dynamics are the focus of this paper we have presented the minimal migration model \eqref{eqn:ode_model} which describes this.
A Python class for accessing and utilising the fields that have been computed for this study is available in a GitHub repository\footnote{\url{github.com/brendanharding/ILFHC_CTD}}.

\section{Vertically symmetric trapezoidal cross-sections}\label{sec:strap}

We begin by studying the dynamics associated with the equilibria for particle migration within trapezoidal cross-sections which are vertically symmetric.
There are a couple of advantages of maintaining vertical symmetry. 
The first, which has practical relevance in the context of applications, is that
any equilibria that occur off the axis of symmetry must exist in pairs having identical radial/lateral coordinate which facilitates the precise collection of focused particles via flow splitting.
The second, which has practical relevance in our implementation of the model, is that we only need to compute the fields in one half of the cross-section and then reflect them appropriately, thus roughly halving the computational cost.

\begin{figure}
\centering
\includegraphics{{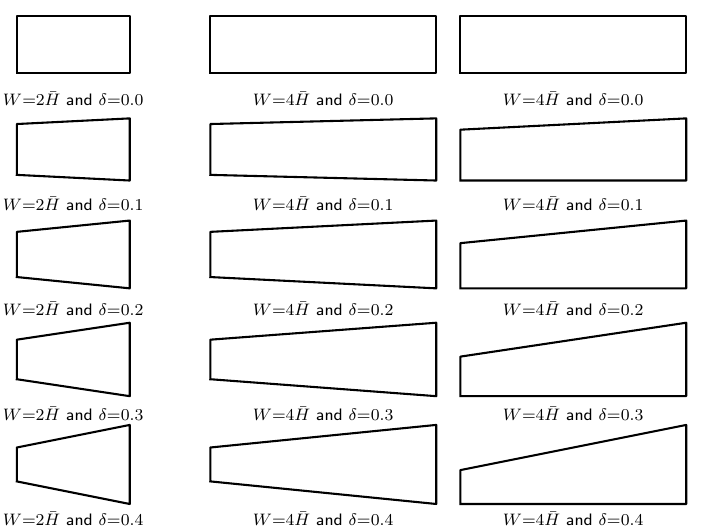}}
\caption{Illustration of the trapezoidal cross-sections considered in this study. 
The left two columns illustrate the vertically symmetric cross-sections considered in Section~\ref{sec:strap}, while the right most column illustrates the asymmetric `flat-bottomed' trapezoidal cross-sections considered in Section~\ref{sec:utrap}.
In each case, $\bar{H}$ is the average duct height, the height at the left wall is $(1-\delta)\bar{H}$ and that at the right wall is $(1+\delta)\bar{H}$.}\label{fig:cs_shapes}
\end{figure}

We study trapezoidal cross-sections with aspect ratios $W/\bar{H}=2,4$ so that a comparison can be made with the rectangular ducts previously studied \cite{HardingStokesBertozzi2019}.
For each aspect ratio, we consider the family of cross-sections $\mathcal{C}_s$, as described in \eqref{eqn:strap_cs}, which are increasingly skewed towards the outside wall with increasing shape parameter $\delta$.
The values of $\delta$ considered in this study are sampled from the interval $[0,0.4]$ (noting some additional results for $\delta=-0.1,0.9$ are provided in appendix \ref{app:extraD}). 
Observe that $\delta=0$ produces curved rectangular ducts (for which the focusing behaviour has been extensively studied).
The specific shape of the cross-sections considered are illustrated in Figure~\ref{fig:cs_shapes}.
The two left most columns show the vertically symmetric families of cross-sections considered in this section, while the right-most column shows the asymmetric cross-sections which will be considered in section \ref{sec:utrap}.

We first examine how the trapezoidal shape influences the background flow.
Then we illustrate how this is also reflected in the features of the $\hat{L}_\ast,\hat{S}_\ast,\hat{D}_\ast$ fields required to compute particle migration.
Following this, we explore the bifurcations that take place as the $\kappa$ parameter is varied for each $\delta$ (and a fixed value of $\alpha$). 
In this way we are able to identify a cusp bifurcation with respect to these two parameters in the case of the larger aspect ratio family.
Lastly, we discuss the differences in migration dynamics that occur with particle size.
In each case we pay special attention to the changes that occur with increasing value of $\delta$.

\subsection{Changes in background flow features}\label{sec:strap_bg_features}

\begin{figure}
\centering
\includegraphics{{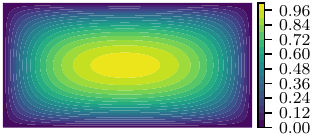}} \qquad
\includegraphics{{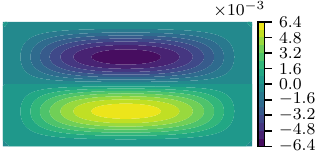}} \\
\includegraphics{{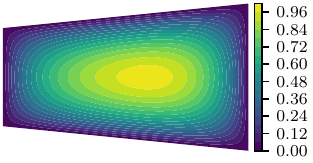}} \qquad
\includegraphics{{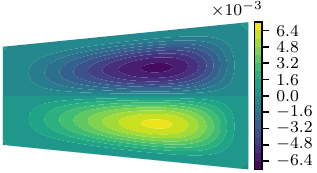}} \\
\includegraphics{{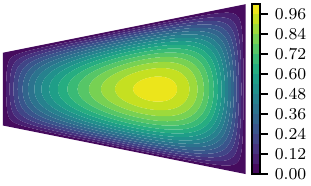}} \qquad
\includegraphics{{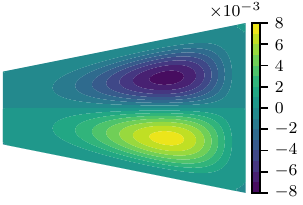}} 
\caption{The change in background flow profile with increasing value of the trapezoidal shape parameter, from top to bottom $\delta=0,0.2,0.4$, for vertically symmetric trapezoids with aspect ratio $W/\bar{H}=2$. 
The left column shows the axial velocity while the right column shows the streamfunction of the secondary flow velocities.
The curvature parameter is $\epsilon=1/80$ and the Dean number is $K=1$.}\label{fig:strap_bg_examples}
\end{figure}

We first illustrate the change with increasing $\delta$ in some properties of the background flow, that is the steady fluid flow in the absence of suspended particles.
The background flow is calculated using a specific scaling of the steady curved duct flow equations which leads to a non-zero streamfunction in the limit of vanishing Dean number \cite{HardingANZIAMJ2019}, consistent with the asymptotic scaling of the secondary flow \cite{Harding2022}.

\begin{figure}
\centering
\includegraphics{{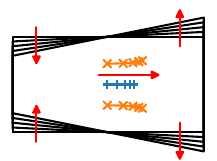}} \qquad
\includegraphics{{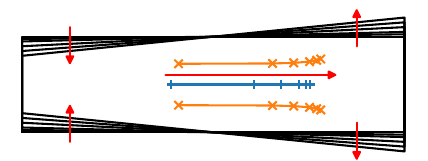}}
\caption{The movement of background flow features with respect to $\delta$ for vertically symmetric trapezoidal cross-sections having aspect ratios $W/\bar{H}=2$ (left) and $W/\bar{H}=4$ (right). 
As $\delta$ increases, the heights of the side walls change as indicated by the vertical red arrows while the horizontal red arrow shows the direction in which the positions of the maximum axial flow velocity (blue $+$) and centre of the secondary flow vortices (orange $\times$) move.
Markers correspond to the values $\delta=0,0.1,0.2,0.3,0.4$ in the $W/\bar{H}=2$ case (left) and $\delta=0,0.05,0.1,0.2,0.3,0.4$ in the $W/\bar{H}=4$ case (right). 
In both cases $\epsilon=1/80$ and the Dean number is $K=1$. 
The outside wall relative to the bend is on the right side.}\label{fig:strap_bg_features}
\end{figure}

Figure~\ref{fig:strap_bg_examples} shows the change in the axial velocity and streamfunction representing the secondary flow motion over a few specific $\delta$ in the case $W/\bar{H}=2$. 
As $\delta$ increases the extrema of these two fields describing $\bar{\vec{u}}$ shift towards the outside wall (right side in the figure).
Figure~\ref{fig:strap_bg_features} summarises the movement of these features for both values of $W/\bar{H}=2,4$ over the range of $\delta$ considered in this study (with fixed values of the curvature parameter $\epsilon=1/80$ and Dean number $K=1$). 

The movement of the location of the maximum axial velocity and the centre of the secondary flow vortices towards the outside wall is primarily due to the cross-sectional area increasing in a neighbourhood of the outside wall while simultaneously decreasing in a neighbourhood of the inside wall. 
This causes the fluid to flow more freely through the duct closer to the outside wall, so that more centrifugal force is generated in this region leading to a similar shift in the secondary flow vortices.
Figure~\ref{fig:strap_bg_features} illustrates that the horizontal shift is most sensitive to $\delta$ around $\delta=0$, particularly in the case of the wider aspect ratio $W/\bar{H}=4$. 
As $\delta$ increases, the magnitude of shift relative to the change in $\delta$ decreases.

These changes in the background flow profile are expected to influence the cross-sectional migration of particles in two fundamental ways. 
Firstly, the shift in features of the secondary flow has a direct impact on the streamlines that small particles would follow upon neglecting all inertial forces.
Secondly, the shift in the axial flow velocity impacts the inertial lift, the force responsible for particle migration across background flow streamlines.
Consequently, we anticipate a similar shift in the locations where particles focus, particularly in the limits of large and small $\kappa$ where the secondary drag and inertial lift force is dominant, respectively.

\subsection{Inertial and drag force fields}

\begin{figure}
\centering
\hfill\begin{subfigure}[b]{0.42\textwidth} 
\includegraphics{{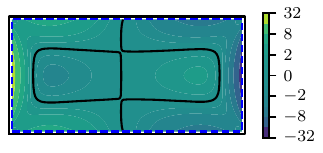}}
\end{subfigure}
\begin{subfigure}[b]{0.42\textwidth} 
\includegraphics{{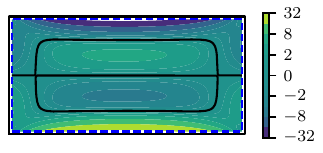}} 
\end{subfigure}\\
\hfill\begin{subfigure}[b]{0.42\textwidth} 
\includegraphics{{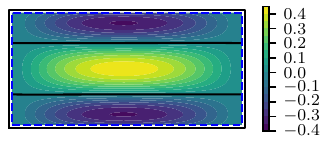}} 
\end{subfigure}
\begin{subfigure}[b]{0.42\textwidth} 
\includegraphics{{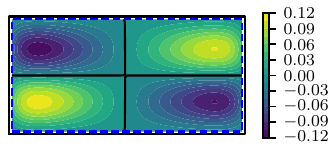}} 
\end{subfigure}\\
\hfill\begin{subfigure}[b]{0.42\textwidth} 
\includegraphics{{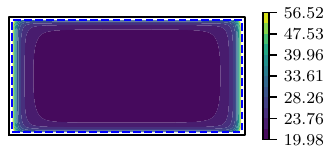}} 
\end{subfigure}
\begin{subfigure}[b]{0.42\textwidth} 
\includegraphics{{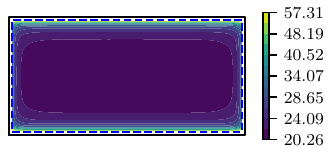}} 
\end{subfigure}
\caption{The fields $\hat{L}_\ast$ (top), $\hat{S}_\ast$ (middle) and $\hat{D}_\ast$ (bottom) for a rectangular cross-section. 
At left are $\hat{L}_r,\hat{S}_r,\hat{D}_r$ and at right are $\hat{L}_z,\hat{S}_z,\hat{D}_z$.
A symmetric log colour scale has been used for the magnitude of the $\hat{L}_\ast$ fields outside of the interval $[-1,1]$.
A log colour scale has been used for the magnitude of the $\hat{D}_\ast$ fields.
The black curves show the zero level sets.
Here we have fixed $\alpha=0.05$, $\epsilon=1/80$, $\delta=0$ and $W/\bar{H}=2$.
The $\hat{D}_\ast$ fields are roughly constant except very close to the walls where they increase sharply, which is barely perceptible.}\label{fig:LSD_2x1_d0}
\end{figure}

In this section we describe some representative examples of the $\hat{L}_\ast,\hat{S}_\ast,\hat{D}_\ast$ fields, where $\ast=r$ or $z$, describing the inertial lift, secondary flow drag and drag coefficient, respectively.
Figure~\ref{fig:LSD_2x1_d0} illustrates these fields in the case of a rectangular cross-section ($\delta=0$) with aspect ratio $W/\bar{H}=2$ and fixed values of $\epsilon=1/80$ and $\alpha=0.05$.
Notice that there is almost, but not exactly, a horizontal symmetry to these results. 
The fields exhibit a very slight skew towards the inside wall relative to the bend (left side) because this provides a shorter path through the curved ducts (and the low flow rate approximation used means the centrifugal force does not overcome this).
The drag coefficient fields are approximately constant through most of the cross-section, with a value a little more than the $6\pi$ obtained from Stokes' drag law, but do increase significantly in a close neighbourhood of the walls.

\begin{figure}
\centering
\hfill\begin{subfigure}[b]{0.42\textwidth} 
\includegraphics{{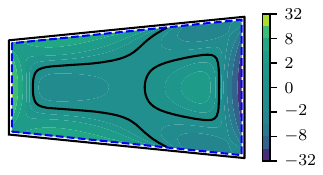}}
\end{subfigure}
\begin{subfigure}[b]{0.42\textwidth} 
\includegraphics{{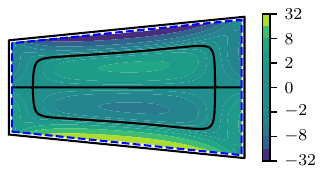}} 
\end{subfigure}\\
\hfill\begin{subfigure}[b]{0.42\textwidth} 
\includegraphics{{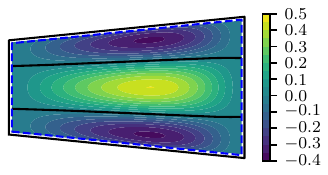}} 
\end{subfigure}
\begin{subfigure}[b]{0.42\textwidth} 
\includegraphics{{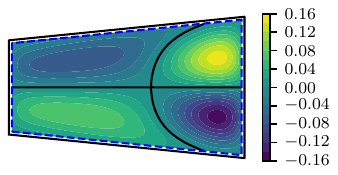}} 
\end{subfigure}\\
\hfill\begin{subfigure}[b]{0.42\textwidth} 
\includegraphics{{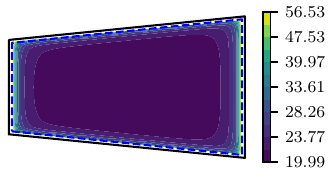}} 
\end{subfigure}
\begin{subfigure}[b]{0.42\textwidth} 
\includegraphics{{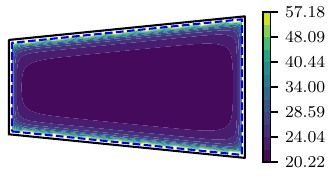}} 
\end{subfigure}
\caption{
The fields $\hat{L}_\ast$ (top), $\hat{S}_\ast$ (middle) and $\hat{D}_\ast$ (bottom) for a trapezoidal cross-section. 
At left are $\hat{L}_r,\hat{S}_r,\hat{D}_r$ and at right are $\hat{L}_z,\hat{S}_z,\hat{D}_z$.
A symmetric log colour scale has been used for the magnitude of the $\hat{L}_\ast$ fields outside of the interval $[-1,1]$.
A log colour scale has been used for the magnitude of the $\hat{D}_\ast$ fields.
The black curves show the zero level sets.
Here we have fixed $\alpha=0.05$, $\epsilon=1/80$, $\delta=0.2$ and $W/\bar{H}=2$.
Compare with the case of a rectangular duct in figure~\ref{fig:LSD_2x1_d0}.}\label{fig:LSD_s2x1_d2}
\end{figure}

Figure~\ref{fig:LSD_s2x1_d2} illustrates the same fields in the case of a trapezoidal cross-section with shape parameter $\delta=0.2$, aspect ratio $W/\bar{H}=2$ and the same fixed values of $\epsilon=1/80$ and $\alpha=0.05$.
We saw in Figure~\ref{fig:strap_bg_examples} how the trapezoidal shape leads to a shift in the axial flow contours and secondary flow streamlines towards the outside wall (right side), and this is directly reflected in changes observed in the fields $\hat{S}_r,\hat{S}_z,\hat{L}_r,\hat{L}_z$.
However, a significant change is seen in the topology of the zero level set of $\hat{L}_r$, which now consists of two non-intersecting curves.
This has important implications for the dynamics noting that previous studies have highlighted the importance of these zero level curves not only in determining the location of equilibria but also in identifying structures that produce migration behaviour indicative of a slow manifold \cite{HardingStokesBertozzi2019}.
The drag coefficients are again approximately constant throughout  the cross-section except in a neighbourhood of the walls.
An illustration of the change in the $\hat{L}_\ast,\hat{S}_\ast$ fields in the case of cross-sections $W/\bar{H}=4$ is deferred to section~\ref{sec:LSD_4x1} when we also compare with flat-bottomed trapezoids. 

\begin{figure}
\centering
\hfill\begin{subfigure}[b]{0.3\textwidth} 
\includegraphics{{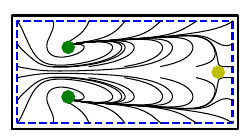}}
\end{subfigure}
\begin{subfigure}[b]{0.6\textwidth} 
\includegraphics{{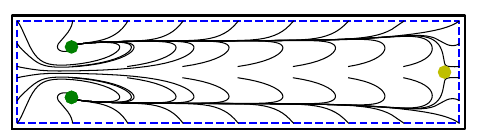}} 
\end{subfigure}\\
\hfill\begin{subfigure}[b]{0.3\textwidth} 
\includegraphics{{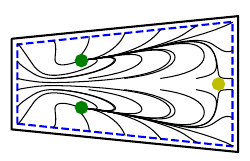}}
\end{subfigure}
\begin{subfigure}[b]{0.6\textwidth} 
\includegraphics{{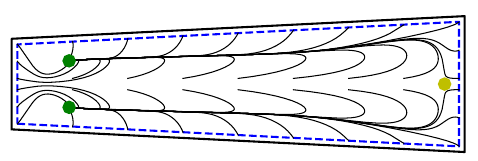}} 
\end{subfigure}\\
\hfill\begin{subfigure}[b]{0.3\textwidth} 
\includegraphics{{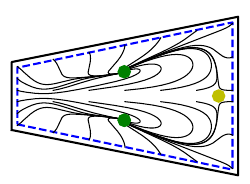}}
\end{subfigure}
\begin{subfigure}[b]{0.6\textwidth} 
\includegraphics{{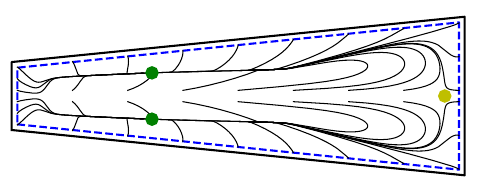}} 
\end{subfigure}
\caption{Particle trajectories for different values of $\delta=0,0.2,0.4$ (top to bottom), aspect ratios $W/\bar{H}=2$ (left) and $W/\bar{H}=4$ (right), and fixed values of $\alpha=0.10$ and $\epsilon=1/160$.
The location of stable equilibria are marked in green, the location of saddle equilibria are marked in yellow.
The marker size reflects the particle size.
The blue dashed line shows the locations at which the particle would touch the cross-section wall.}\label{fig:strap_traj}
\end{figure}

Figure~\ref{fig:strap_traj} illustrates the cross-sectional trajectories for a single particle of size $\alpha=0.1$ from different starting locations in a trapezoidal cross-section for several $\delta$ values, the two aspect ratios $W/\bar{H}=2,4$ and fixed $\epsilon^{-1}=160$. 
The change in the trajectories with increasing $\delta$ is generally quite subtle.
In each figure we observe that particles released from different initial positions initially focus onto one of two curves.
Particles subsequently migrate along these curves towards a stable equilibria, often much slower than the initial migration and thus these curves are sometimes referred to as slow manifolds.
To the right of the stable equilibria, the slow manifolds lie on the heteroclinic orbits connecting the saddle equilibrium to each stable equilibria. 
In the case of $\delta=0.4$ this curve appears to extend to the left of the stable equilibria, although there is no heteroclinic orbit here.
In each plot we observe that the slow manifolds appear to develop a kink as $\delta$ increases, that is a point at which the slope of each curve changes sharply.

\subsection{Bifurcations in the dynamics of a small particle over a large range of bend radii}\label{sec:strap_bifur}

We now examine the change in particle dynamics with changes in the curvature parameter $\epsilon$ and the associated change in the equilibria for several different values of the trapezoidal shape parameter $\delta$.
For this purpose, for each value of $\delta$, we compute estimates of the fields $\hat{L}_\ast$, $\hat{S}_\ast$ and $\hat{D}_\ast$ for a small particle $\alpha=0.05$ and the curvature parameter $\epsilon=1/160$.
Because there is only weak dependence of these fields on $\epsilon$, the one sample at $\epsilon=1/160$ provides a reasonably good approximation for all values $0\leq \epsilon\leq 10^{-2}$ and we are able to capture the particle migration dynamics for different $\epsilon$ (in this range) by changing the value of $\kappa=4\epsilon/\alpha^3=32000\epsilon$ in \eqref{eqn:ode_model}.
Small changes in $\kappa$ (in relative terms) might also be interpreted as due to small changes in $\alpha$ while holding $\epsilon$ fixed, but we'll return to this point later.
Given our estimates of the $\hat{L}_\ast,\hat{S}_\ast,\hat{D}_\ast$ fields, we proceed to find and classify all fixed points in the cross-section over many values of $\kappa\in[0.5,200]$, which is an interval over which most bifurcations occur. 
For convenience we describe results with cross-sectional coordinates scaled so that $\bar{H}=2$.

\begin{figure}
\centering
\includegraphics{{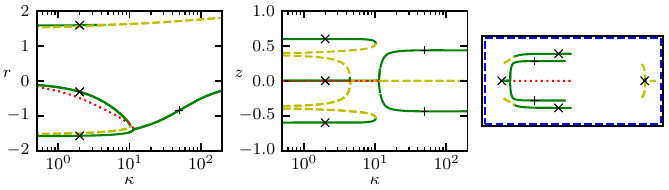}}
\includegraphics{{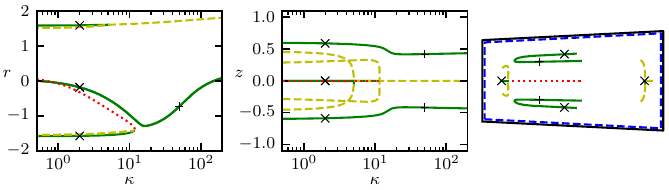}}
\includegraphics{{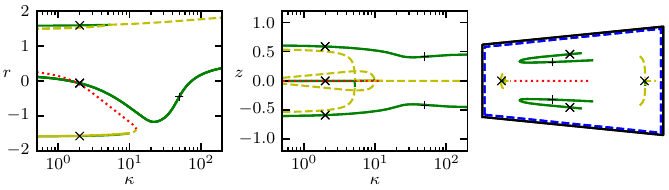}}
\includegraphics{{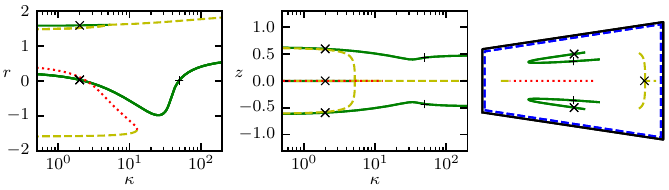}}
\includegraphics{{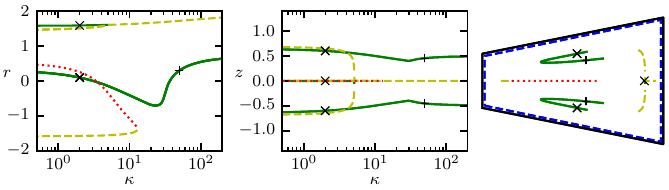}}
\caption{Dynamics associated with the fixed points of \eqref{eqn:ode_model} for particle size $\alpha=0.05$, aspect ratio $W/\bar{H}=2$, and trapezoidal shape parameter $\delta=0,0.1,0.2,0.3,0.4$ (top to bottom). 
Line styles denote stability: green solid for stable, yellow dashed for saddle, red dotted for unstable.
The left and centre column show the horizontal and vertical coordinate, respectively, of fixed points vs $\kappa$ (or equivalently $32000\epsilon$)
The right column shows the path followed by fixed points within the cross-section as $\kappa$ changes. 
The cross and plus markers illustrate the location of stable fixed points at the specific values of $\kappa=2,50$, respectively.
Cross-sectional coordinates have been re-scaled so that $\bar{H}=2$.}\label{fig:strap_2x1_dynamics}
\end{figure}

We begin by examining the family of vertically symmetric trapezoidal cross-sections having aspect ratio $W/\bar{H}=2$. 
Figure~\ref{fig:strap_2x1_dynamics} illustrates the bifurcations that occur with respect to $\kappa$ for each value of the shape parameter $\delta=0,0.1,0.2,0.3,0.4$. 
In the context of particle sorting/separation, our main interest is in what happens to the stable fixed points, and specifically the stable pair towards which most particles focus.
Thus, our discussion focuses on the bifurcations that influence stable fixed points and how these change as $\delta$ increases.

We start with a brief description of the bifurcations that occur in the case $\delta=0$, which has already been explored in detail \cite{valani2021_b}.
Notice there is a (vertically symmetric) pair of stable equilibria located near $(r,z)\approx(0,\pm0.6)$ for small $\kappa$.
As $\kappa$ increases these stable equilibria migrate towards the inside wall (left side) and at $\kappa\approx10$ each meets a saddle equilibria and a saddle node bifurcation occurs.
Around the same time a supercritical pitchfork bifurcation occurs at a stable equilibrium which has been located near $(r,z)\approx(-1.5,0)$ thus far. 
A pair of stable equilibria produced in this bifurcation quickly divert away from $z=0$ towards $z\approx\pm 0.4$ between $\kappa=10$ and $\kappa=20$.
As $\kappa$ increases towards $100$ this stable equilibrium pair moves laterally away from the inside wall towards the centre line $r=0$.
For small $\kappa$ there is also a stable equilibrium located near $(r,z)=(1.6,0)$ which, in a subcritical pitchfork bifurcation, merges with a pair of saddle equilibria nearby (occurring around $\kappa\approx5$) leaving behind a saddle equilibrium.

Now, in the case $\delta=0.1$ there are several qualitative changes. 
First, the stable equilibrium pair initially located around $(r,z)\approx(0,\pm0.6)$ for $\kappa=1$ migrates towards the inside wall as $\kappa$ increases towards $\kappa\approx10$ but does not meet any saddle equilibria. 
Instead, it migrates back towards $r=0$ as $\kappa$ increases towards $100$ (and beyond).
Notice the $z$ coordinate is a little closer to the origin during the migration back towards $r=0$ for large $\kappa$.
The stable equilibrium initially located near $(r,z)=(-1.6,0)$, for small $\kappa$, remains there as $\kappa$ increases before encountering a pair of saddle equilibria at $\kappa\approx 10$ leaving behind a saddle equilibrium in a subcritical pitchfork bifurcation. 
Almost immediately the saddle equilibrium encounters an unstable equilibrium and the two disappear in a saddle node bifurcation.
The stable equilibrium initially located near $(r,z)=(1.6,0)$ behaves similarly to the $\delta=0$ case.
Another important observation is that the rate at which the $r$ coordinate of the stable equilibrium pair increases (with respect to the logarithmic $\kappa$ scale), after achieving its minimum, becomes faster as $\delta$ increases.

With further increases in $\delta$ the sequence of bifurcations and qualitative behaviour is similar to the $\delta=0.1$ case excepting for two significant changes.
First, for $\delta=0.3,0.4$ we no longer observe a stable equilibrium, nor a nearby pair of saddle equilibria, located near $(r,z)=(-1.6,0)$ for small $\kappa$ values.
This leaves only the one saddle equilibrium located near $(r,z)=(-1.6,0)$ which encounters an unstable equilibrium around $\kappa\approx 10$ causing both to disappear in a saddle node bifurcation.
Second, the main pair of stable equilibria follows a path that, itself, shifts towards the outside wall (right side) of the duct as $\delta$ increases.
The range of $r\in[-2,2]$ covered by this path also decreases for the two samples $\delta>0.2$ which potentially reduces the ability to efficiently separate this particle size from others.
In this context, the value $\delta=0.2$ appears to be close to optimal in the sense of having the largest range of $r$-values at which the stable equilibrium pair can be focused.

\begin{figure}
\centering
\includegraphics{{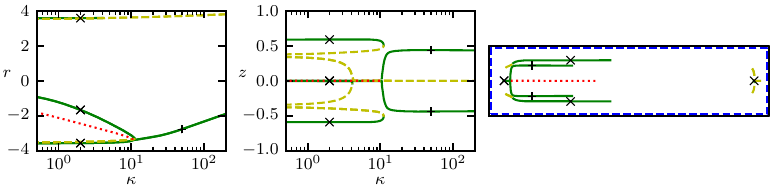}}
\includegraphics{{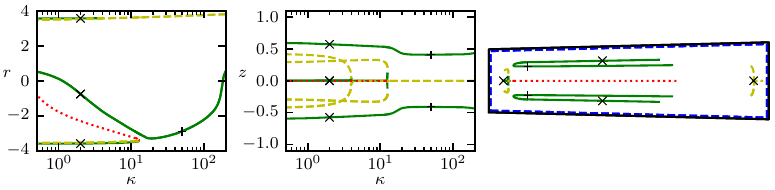}}
\includegraphics{{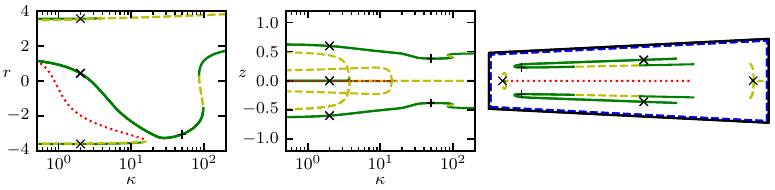}}
\includegraphics{{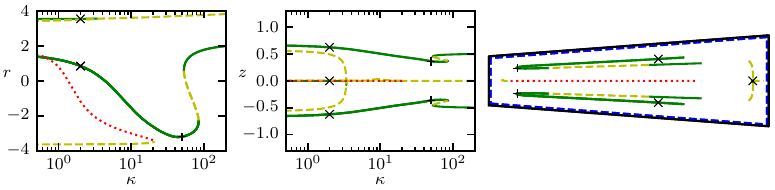}}
\includegraphics{{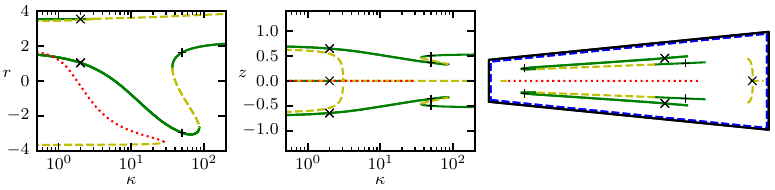}}
\caption{Dynamics associated with the fixed points of \eqref{eqn:ode_model} for particle size $\alpha=0.05$, aspect ratio $W/\bar{H}=4$, and trapezoidal shape parameter $\delta=0,0.1,0.2,0.3,0.4$ (top to bottom). 
Line styles denote stability: green solid for stable, yellow dashed for saddle, red dotted for unstable.
The left and centre column show the horizontal and vertical coordinate, respectively, of fixed points vs $\kappa$ (or equivalently $32000\epsilon$)
The right column shows the path followed by fixed points within the cross-section as $\kappa$ changes. 
The cross and plus markers illustrate the location of stable fixed points at the specific values of $\kappa=2,50$, respectively.
Cross-sectional coordinates have been re-scaled so that $\bar{H}=2$.}\label{fig:strap_4x1_dynamics}
\end{figure}

We now examine the change in bifurcations that occur amongst the family of vertically symmetric trapezoidal cross-sections having aspect ratio $W/\bar{H}=4$. 
The same methodology as was used in the $W/\bar{H}=2$ case is adopted here, that is the fields are estimated for fixed $\epsilon=1/160$ and $\alpha=0.05$ and we then explore the effect of changing $\kappa$ in \eqref{eqn:ode_model}.
Figure~\ref{fig:strap_4x1_dynamics} illustrates the fixed point dynamics for the different values of $\delta=0,0.1,0.2,0.3,0.4$.
Again, our main interest is the movement of stable fixed points. 

The general dynamics in the cases of $\delta=0$ and $\delta=0.1$ are qualitatively the same as described previously in the case of the smaller aspect ratio. 
One interesting quantitative difference is the significant increase in the horizontal range of the stable equilibrium pair on increasing $\delta$ form $0$ to $0.1$.
This is not unexpected because of the similar large movement in the features of the background flow discussed in section \ref{sec:strap_bg_features}.
Additionally, observe that for $\delta=0.1$ and $\kappa$ between $100$ and $200$ there is a relatively sharp change in the $r$ coordinate of the stable equilibrium pair (with respect to the logarithmic $\kappa$ scale).
For larger values of $\kappa$ (outside the range shown) the $r$ vs $\kappa$ curve flattens out and approaches the location of the centre of the Dean vortices ($r\approx 1.75$).

As $\delta$ increases beyond $0.1$, the most significant change is the development of a fold in the $r$ vs $\kappa$ curve near to $\kappa=100$ (this also presents in the $z$ vs $\kappa$ curve but is a little more difficult to see).
As $\delta$ increases the range of $r$ values covered by the fold increases.
For $\delta=0.4$ the fold covers values of $\kappa$ approximately spanning $40$ to $100$.
If one were to fix a value of $\kappa$ in the (approximate) range $40$ to $100$ and examine the behaviour of the system as $\delta$ increases from $0$ to $0.4$, then the `fold' would generally present as a saddle node bifurcation (occurring somewhere in the range $r\in[0,1]$). 
The one exception is at the specific value of $\kappa$ where the fold first occurs which possibly presents as a pitchfork bifurcation (followed soon after by a saddle node bifurcation involving the left most stable equilibrium pair).

The development of the fold due to the cusp bifurcation has important implications on applications involving size-based particle separation.
Over the range of $\kappa$ covered by the fold there exists two (vertically symmetric) stable equilibria pairs separated by a saddle equilibria pair.
Having stable equilibria for one particle size at two separated radial locations potentially inhibits the ability to efficiently separate those particles from others suspended in the flow.
Of course, the basins of the two stable equilibrium pairs needs to be examined in more detail.
The use of sheath flow or pre-focusing regions could eliminate this issue by facilitating the selection of one of the two stable equilibria pairs in this parameter range. 

Observe that over $0.2\leq \delta\leq 0.4$ a stable equilibria pair is reasonably close to the inside wall (relative to the width of the cross-section) for a value of $\kappa$ between $10$ and $100$, unlike the case of the smaller aspect ratio cross-section where there is a noticeable increase in the minimum $r$ value over the same range of $\delta$.
Lastly we note that the existence and dynamics of stable and saddle equilibria near $(r,z)\approx(\pm3.5,0)$ for small values of $\kappa$ are similar to that observed for the smaller aspect ratio cross-sections.

In this section we have examined bifurcations based on changing $\kappa$ in \eqref{eqn:ode_model} using estimates of the $\hat{L}_\ast,\hat{S}_\ast,\hat{D}_\ast$ fields obtained with the specific values $\epsilon=1/160$ and $\alpha=0.05$.
As discussed, given the weak dependence of the fields on $\epsilon$ over the range $\epsilon\leq 1/100$, we generally interpret changes in $\kappa=4\epsilon/\alpha^3$ as being due to changes in $\epsilon$ with a fixed $\alpha$.
However, for small changes in $\alpha$, e.g. on the order of $10\%$, we also expect very little change in the fields $\hat{L}_\ast$, $\hat{S}_\ast$ and $\hat{D}_\ast$.
Because of the inverse cubic scaling of $\alpha$'s contribution to $\kappa$, a $10\%$ decrease in $\alpha=0.05$ increases $\kappa$ by $37\%$, and a $10\%$ increase in $\alpha$ decreases $\kappa$ by $25\%$.
Therefore, changes in $\kappa$ of this magnitude can alternatively be interpreted as being due to small changes in $\alpha$ with fixed $\epsilon$.
Then, within the curves describing the $r$ coordinate of stable equilibria as a function of $\kappa$, the presence of a rapid change provides an opportunity to separate particles having a relatively small difference in size.
The onset of the cusp bifurcation, occurring somewhere in the interval $0.1<\delta<0.2$, is a specific location where this might be exploited to great benefit.
For larger changes in $\alpha$ there are subtle changes in the $\hat{L}_\ast,\hat{S}_\ast,\hat{D}_\ast$ fields that become important and, as such, we compare curves describing the horizontal location of the stable equilibria pair for $\alpha=0.05,0.10,0.15,0.20$ in the following section.

\subsection{Comparison of dynamics for different particle sizes}\label{sec:strap_alpha_effects}

\begin{figure}
\centering
\includegraphics{{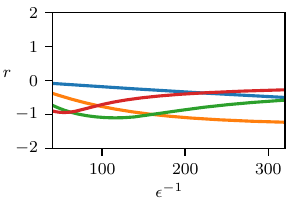}} \,
\includegraphics{{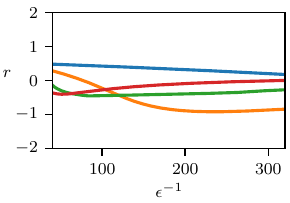}} \,
\includegraphics{{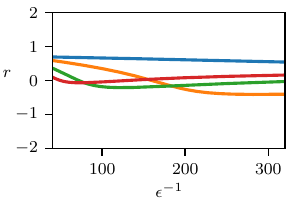}} \\
\includegraphics{{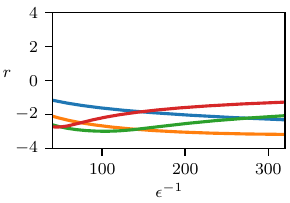}} \,
\includegraphics{{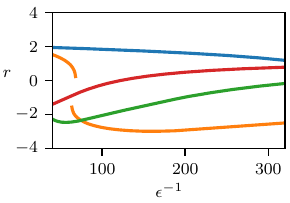}} \,
\includegraphics{{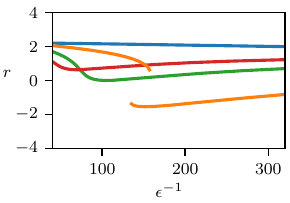}} \\
\caption{The horizontal location of vertically symmetric stable equilibrium pairs versus $\epsilon^{-1}$ for $\delta=0,0.2,0.4$ (left to right) and aspect ratios $W/\bar{H}=2,4$ (top, bottom). 
The colour of the curve corresponds to particle sizes $\alpha=0.05$ (blue), $\alpha=0.1$ (orange), $\alpha=0.15$ (green), $\alpha=0.2$ (red).}\label{fig:strap_eps_trends}
\end{figure}

For the results of this section we produced estimates of the $\hat{L}_\ast,\hat{S}_\ast,\hat{D}_\ast$ fields for each combination of the four values $\alpha=0.05,0.10,0.15,0.20$, the two values $\epsilon=1/160,1/80$ and the three values $\delta=0,0.2,0.4$.
We use this data to focus on two key questions relating to the horizontal location of stable equilibrium pairs.
First, for each of the three $\delta$ values, if we interpolate/extrapolate the fields obtained from the two distinct $\epsilon$ to examine a practical range of duct bend radii, specifically $\epsilon^{-1}\in[40,320]$, what degree of separation is achieved among the four particle sizes?
Second, for each of the three $\delta$ values, if we take the $\epsilon=1/160$ samples and examine the change in horizontal location of stable equilibrium pairs by varying $\kappa\in[0.5,200]$ in \eqref{eqn:ode_model}, i.e. similar to what was done in section \ref{sec:strap_bifur}, what differences are observed in the trends amongst the four particle sizes?

\begin{figure}
\centering
\includegraphics{{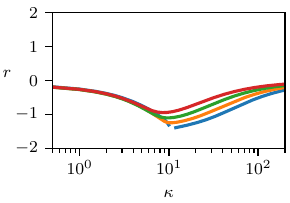}} \,
\includegraphics{{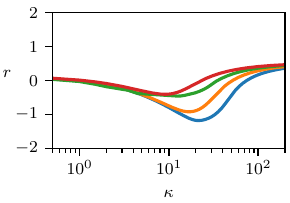}} \,
\includegraphics{{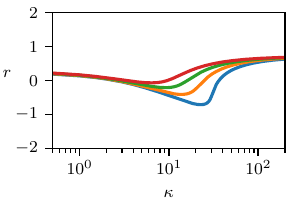}} \\
\includegraphics{{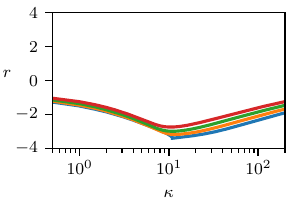}} \,
\includegraphics{{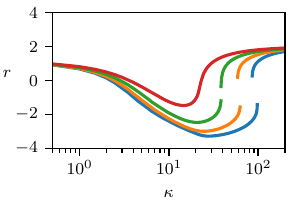}} \,
\includegraphics{{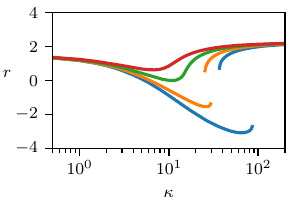}} \\
\caption{Change in horizontal location of the stable equilibrium pair with respect to $\kappa$ for $\delta=0,0.2,0.4$ (left to right) and aspect ratios $W/\bar{H}=2,4$ (top, bottom). 
The colour of the curve corresponds to particle sizes $\alpha=0.05$ (blue), $\alpha=0.1$ (orange), $\alpha=0.15$ (green), $\alpha=0.2$ (red).}\label{fig:strap_kap_trends}
\end{figure}

Figure~\ref{fig:strap_eps_trends} addresses the first of these questions.
It shows the horizontal location of the stable equilibrium pair over a practical range of $\epsilon^{-1}$.
A general application of these plots is to facilitate the selection of an appropriate bend radius to achieve good horizontal separation between specific particle sizes.
The top row shows the result for the $W/\bar{H}=2$ cross-sections for each of $\delta=0,0.2,0.4$, left to right, with each curve denoting a different particle size (with $\alpha=0.05,0.1,0.15,0.2$ coloured blue, orange, green, and red, respectively).
The bottom row shows a similar result but for the $W/\bar{H}=4$ cross-sections.

We see a general trend that the curves shift upwards with increasing $\delta$.
Delving deeper, for $\delta=0$ (the rectangular duct case) no pair of particle sizes is separated by more than approximately $W/4$, but there are some opportunities to achieve a small separation of one particle size from the other three (e.g. the $\alpha=0.05$ particle for $\epsilon^{-1}<100$ and the $\alpha=0.1$ particle for $\epsilon^{-1}>200$ within both aspect ratios).
For $\delta=0.2$ we see a significant change.
In general terms, much greater separation between the different particle sizes can be achieved and there is a large range of $\epsilon^{-1}$ over which the ordering remains consistent and all four particles achieve separation from one another.
There is an especially large degree of separation in the results for the wider duct.
Notice that for $W/\bar{H}=4$, a particle of size $\alpha=0.1$ is affected by a cusp bifurcation within this range of $\epsilon$.
For $\delta=0.4$ similar observations can be made in comparison to the $\delta=0$ case, although it is clear the degree of separation is reduced somewhat compared to the $\delta=0.2$ case.
The two parts of the fold that develops from a cusp bifurcation involving the $\alpha=0.1$ particle remains evident but has shifted to larger $\epsilon^{-1}$ values.

Figure~\ref{fig:strap_kap_trends} addresses the second question by showing how the horizontal location of stable fixed point pairs change versus $\kappa$ (remembering from section \ref{sec:strap_bifur} that changes in $\kappa$ should generally be interpreted as due to changes in $\epsilon$ in this context).
The top row shows the result for the $W/\bar{H}=2$ cross-sections for each of $\delta=0,0.2,0.4$, left to right, with each curve denoting a different particle size (with $\alpha=0.05,0.1,0.15,0.2$ coloured blue, orange, green, and red, respectively).
It should be noted that for the largest two particle sizes, the larger values of $\kappa$ shown in these plots correspond to bend radii at which our model may be less accurate (i.e. corresponding to $\epsilon>1/100$), but we have included these results to provide a comparison over the entire $\kappa$ range shown.

The left most plots of figure~\ref{fig:strap_kap_trends} show the approximate collapse in the four curves previously observed for rectangular ducts \cite{HardingStokesBertozzi2019}.
As $\delta$ increases (left to right columns), the four curves diverge increasingly in the middle of the $\kappa$ range but come together at each end.
The plots for $W/\bar{H}=4$ also illustrate the subtle way in which the development of the fold with increasing $\delta$ changes with the different values of $\alpha$.
Near to the onset of the folds, at $\delta=0.2$, we see that the value of $\kappa$ around which these occur decreases with increasing particle size.
Moreover, for the largest particle size ($\alpha=0.2$) considered in this study there is no fold at all.
For $\alpha=0.15$ there is a small fold present for $\delta=0.2$ but none for $\delta=0.4$, indicating a second cusp bifurcation has occurred.
For the two smallest particles, the range of $\kappa$ over which the fold exists is larger for $\delta=0.4$ than for $\delta=0.2$, and is also larger for $\alpha=0.05$ than it is for $\alpha=0.10$ for both $\delta=0.2,0.4$.
Another general trend to note is the minimum value of $r$ achieved in each case increases with particle size.

\section{Flat-bottomed trapezoidal cross-sections}\label{sec:utrap}

We now turn our attention to examining particle migration within the flat-bottomed family of trapezoidal cross-sections.
The vertically asymmetric family of cross-sections is denoted by $\mathcal{C}_a$, defined in \eqref{eqn:utrap_cs}.
For brevity, we consider a single aspect ratio $W/\bar{H}=4$ within this family. 
The specific cross-section shapes we examine are illustrated in the right column of figure~\ref{fig:cs_shapes}.
The sub-sections that follow have the same structure as our study of the vertically symmetric family of cross-sections in section~\ref{sec:strap}.

\subsection{Changes in background flow features}\label{sec:utrap_bg_features}

\begin{figure}
\centering
\includegraphics{{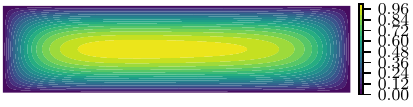}} \quad
\includegraphics{{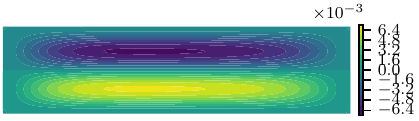}} \\
\includegraphics{{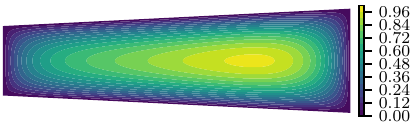}} \quad
\includegraphics{{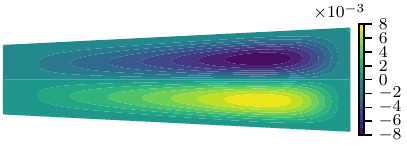}}  \\
\includegraphics{{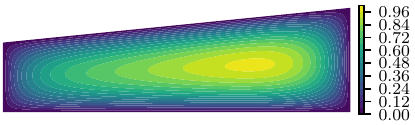}} \quad
\includegraphics{{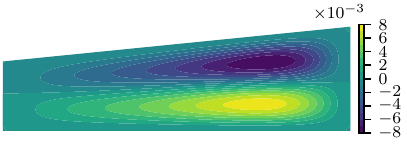}} 
\caption{
Background flow profiles for several cross-section shapes having common aspect ratio $W/\bar{H}=4$. 
The left column shows the axial velocity and the right column shows the streamfunction of the secondary flow.
The cross-sections are rectangular (top row), vertically symmetric trapezoid with $\delta=0.2$ (middle row) and flat-bottomed trapezoid with $\delta=0.2$ (bottom row).
In each case, the curvature parameter is $\epsilon=1/80$ and the Dean number is $K=1$.
}\label{fig:utrap_bg_examples}
\end{figure}

\begin{figure}
\centering
\includegraphics{{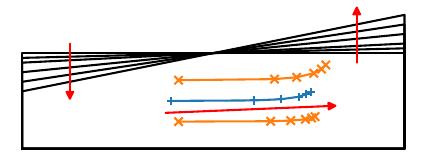}}\qquad
\includegraphics{{figures_pdf/strap_4x1_R80_K1_bgf_features-eps-converted-to.pdf}}
\caption{
The movement of background flow features with respect to $\delta$ for flat-bottomed (left) and vertically symmetric (right) trapezoidal cross-sections having aspect ratio $W/\bar{H}=4$. 
As $\delta$ increases, the heights of the side walls change as indicated by the vertical red arrows, while the horizontal red arrow shows the direction in which the positions of the maximum axial flow velocity (blue $+$) and centre of the secondary flow vortices (orange $\times$) move.
Markers correspond to the values $\delta=0,0.05,0.1,0.2,0.3,0.4$. 
The curvature parameter is $\epsilon=1/80$ and the Dean number is $K=1$. 
The outside wall relative to the bend is on the right side.
}\label{fig:utrap_bg_features}
\end{figure}

Analogous to section~\ref{sec:strap_bg_features} we examine the change in background flow features with increasing $\delta$.
Figure~\ref{fig:utrap_bg_examples} shows both the axial fluid velocity and the streamfunction of the secondary fluid flow through curved ducts having a rectangular, vertically symmetric trapezoidal and flat-bottomed trapezoidal cross-section, each with aspect ratio $W/\bar{H}=4$.
Figure~\ref{fig:utrap_bg_features} summarises the change in position of the maximum axial velocity and centres of the secondary flow vortices, with respect to $\delta$, for ease of comparison the vertically symmetric cross-section is shown together with the flat-bottomed trapezoidal cross-section.
The effect of the trapezoidal shape is qualitatively similar for the vertically symmetric and flat-bottomed cases, except for some relatively small differences due to the presence/absence of vertical symmetry.
In particular, as for the symmetric trapezoid, the asymmetric trapezoidal shape causes the location of the maximum axial flow velocity and the centres of the secondary flow vortices to shift toward the outer wall relative to their positions in the rectangular duct. 
The absence of vertical symmetry in the flat-bottomed cross-section leads to the centre of the upper vortex being located slightly to the right of that of the lower vortex. 

We see that even a slight trapezoidal shape ($\delta=0.05$) has a significant effect in pushing the features towards the outside wall and increasing $\delta$ through $0.1,0.2,0.3,0.4$ further pushes features toward the outer wall, albeit with diminishing effect.
However, the lack of vertical symmetry in the flat-bottomed trapezoidal cross-section results in an upward movement of the position of maximum axial flow velocity and the centres of both upper and lower secondary flow vortices, with the amount increasing with $\delta$.
This especially noticeable for the location of the upper vortex centre for $\delta\geq 0.2$.
The sub-sections that follow will further illustrate how the absence of vertical symmetry in the flat-bottomed trapezoids influences the dynamics of particle migration.

\subsection{Inertial lift and secondary drag force fields}\label{sec:LSD_4x1}

\begin{figure}
\centering
\hfill\begin{subfigure}[b]{0.48\textwidth} 
\includegraphics{{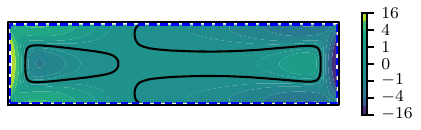}}
\end{subfigure}
\begin{subfigure}[b]{0.48\textwidth} 
\includegraphics{{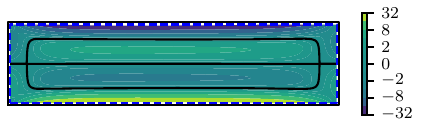}} 
\end{subfigure}\\
\hfill\begin{subfigure}[b]{0.48\textwidth} 
\includegraphics{{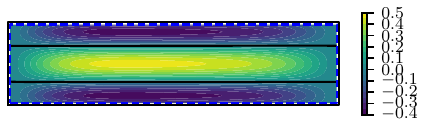}} 
\end{subfigure}
\begin{subfigure}[b]{0.48\textwidth} 
\includegraphics{{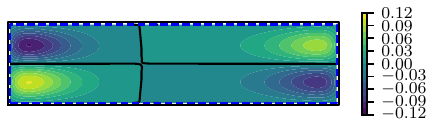}} 
\end{subfigure}
\hrule
\hfill\begin{subfigure}[b]{0.48\textwidth} 
\includegraphics{{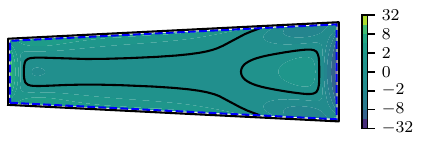}}
\end{subfigure}
\begin{subfigure}[b]{0.48\textwidth} 
\includegraphics{{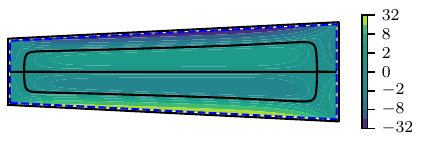}} 
\end{subfigure}\\
\hfill\begin{subfigure}[b]{0.48\textwidth} 
\includegraphics{{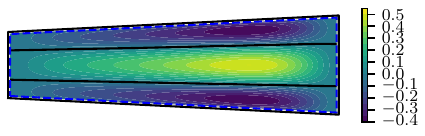}} 
\end{subfigure}
\begin{subfigure}[b]{0.48\textwidth} 
\includegraphics{{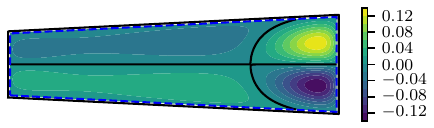}} 
\end{subfigure}
\hrule
\hfill\begin{subfigure}[b]{0.48\textwidth} 
\includegraphics{{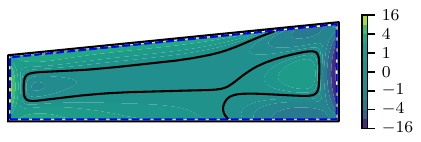}}
\end{subfigure}
\begin{subfigure}[b]{0.48\textwidth} 
\includegraphics{{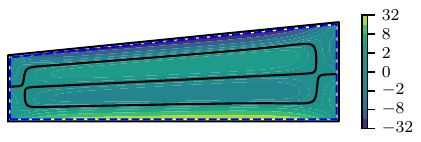}} 
\end{subfigure}\\
\hfill\begin{subfigure}[b]{0.48\textwidth} 
\includegraphics{{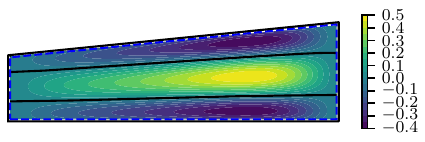}} 
\end{subfigure}
\begin{subfigure}[b]{0.48\textwidth} 
\includegraphics{{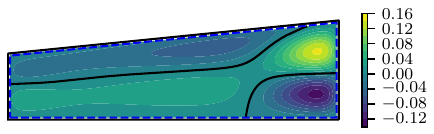}} 
\end{subfigure}
\caption{The fields $\hat{L}_\ast$ (top in each group) and $\hat{S}_\ast$ (bottom in each group), $\ast=r,z$, for a rectangular cross-section (top group), vertically symmetric cross-section having $\delta=0.2$ (middle group) and a flat-bottomed trapezoidal cross-section having $\delta=0.2$ (bottom group), with the left column being the $r$ components and the right column being the $z$ components.
Observe the symmetric log scale in the magnitude of the $\hat{L}_\ast$ fields (outside of the interval $[-1,1]$).
The black curves illustrate the zero level sets.
In each case we have fixed $\alpha=0.05$, $\epsilon=1/80$ and $W/\bar{H}=4$.
}\label{fig:LSD_4x1_d2_all}
\end{figure}

Figure~\ref{fig:LSD_4x1_d2_all} shows the $\hat{L}_\ast$ and $\hat{S}_\ast$ fields, $\ast=r,z$, within several cross-sections having aspect ratio $W/\bar{H}=4$, curvature parameter $\epsilon=1/80$ and particle size $\alpha=0.05$. 
The top group of four shows a rectangular cross-section, the middle group shows a vertically symmetric cross-section with $\delta=0.2$ and the bottom group shows a flat-bottomed cross-section with $\delta=0.2$.
The $\hat{D}_\ast$ fields have been omitted because, as previously noted, they are approximately constant over most of the cross-section (the exception being within a small neighbourhood of the walls, which has limited effect on migration dynamics).

For the rectangular cross-section, the $\hat{L}_\ast,\hat{S}_\ast$ fields show a small skew in features toward the inside wall, similar to that observed in the background flow in figure~\ref{fig:utrap_bg_examples}.
In the trapezoidal cases we see the features shift significantly towards the outside wall, again mirroring what occurs in the background flow.
Notice the differences in topology of the zero level sets associated with the $L_r$ field between each cross-section.
In the rectangular case ($\delta=0$ in either family) there are two disconnected regions where $L_r<0$, and one where $L_r>0$. 
However, in the vertically symmetric trapezoidal cross-section this situation is reversed, and in the flat-bottomed trapezoidal cross-section there is only one of each such region.
Similarly, for the $L_z$ field we observe two regions where $L_z<0$ and another two where $L_z>0$ within both the rectangular and vertically symmetric trapezoidal cross-sections, but only one region for each sign within the flat-bottomed trapezoidal cross-section.
Lastly, another interesting topology change is evident in the zero level set of $S_z$, noting that in the flat-bottomed case there is a single connected region where $S_z>0$.

\begin{figure}

\quad \includegraphics[width=0.45\textwidth]{{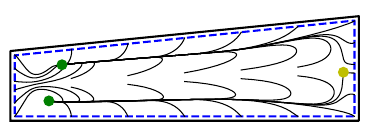}} 
\quad \includegraphics[width=0.45\textwidth]{{figures_pdf/strap_4x1_d2_R160_a10_traj-eps-converted-to.pdf}} 

\quad \includegraphics[width=0.45\textwidth]{{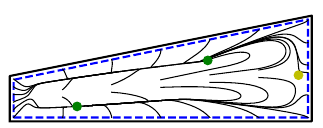}}  
\quad \includegraphics[width=0.45\textwidth]{{figures_pdf/strap_4x1_d4_R160_a10_traj-eps-converted-to.pdf}}

\caption{Particle trajectories with different values of $\delta=0.2,0.4$ (top to bottom) for the families of flat-bottomed (left) and vertically symmetric (right) cross-sections, each with aspect ratio $W/\bar{H}=4$, particle size $\alpha=0.10$ and curvature parameter $\epsilon=1/160$.
The location of stable equilibria are marked in green, the location of saddle equilibria are marked in yellow.
The marker size reflects the particle size.
The blue dashed line shows the locations at which the particle would touch the cross-section wall.}\label{fig:utrap_traj}
\end{figure}

Figure \ref{fig:utrap_traj} compares particle trajectories within flat-bottomed and vertically symmetric cross-sections.
For $\delta=0.2$ (top row) we observe that in both cross-sections the trajectories first migrate onto a slow manifold before migrating along it towards a stable equilibria.
The asymmetry of the flat-bottomed cross-section is mostly evident in the small offset in horizontal location of the two stable equilibria.
For $\delta=0.4$ (bottom row) there are still some broad qualitative similarities in the trajectories within both cross-sections, but we observe a significant difference in the horizontal location of the two stable equilibria in the flat-bottomed case.
This is atypical across the broader parameter space, which will be explored in the following sections, but can occur.

\subsection{Bifurcations in the dynamics of a small particle over a large range of bend radii}

\begin{figure}
\centering
\includegraphics{{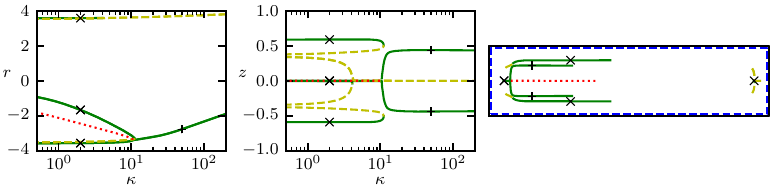}}
\includegraphics{{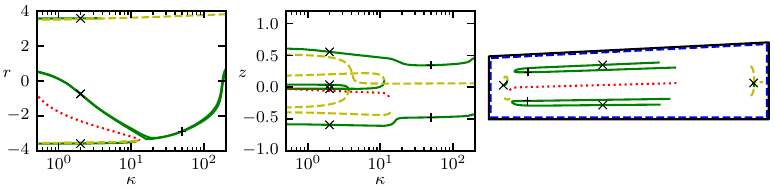}}
\includegraphics{{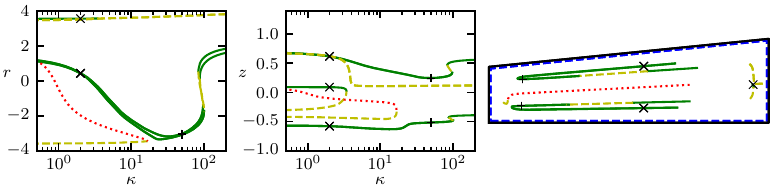}}
\includegraphics{{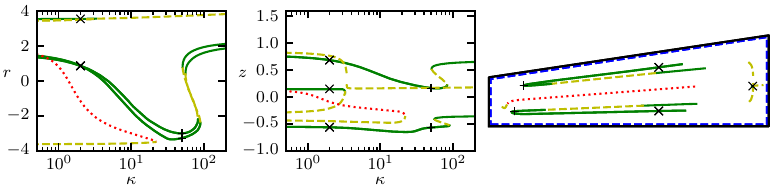}}
\includegraphics{{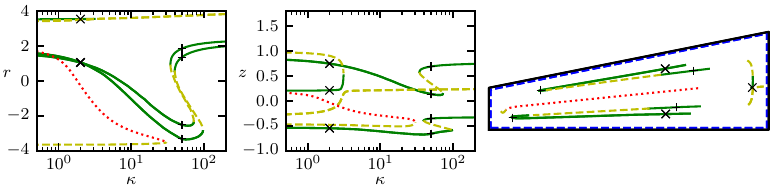}}
\caption{Dynamics associated with the fixed points of \eqref{eqn:ode_model} for particle size $\alpha=0.05$, aspect ratio $W/\bar{H}=4$, and flat-bottomed trapezoidal shape parameters $\delta=0,0.1,0.2,0.3,0.4$ (top to bottom). 
Line styles denote stability: green solid for stable, yellow dashed for saddle, red dotted for unstable.
The left and centre columns show the horizontal and vertical coordinate, respectively, of equilibria vs $\kappa$ (or equivalently $32000\epsilon$). 
The right column shows the paths of equilibria within the cross-section.
The cross and plus markers illustrate the location of stable equilibria at the specific values of $\kappa=2,50$, respectively.
Cross-sectional coordinates have been re-scaled so that $\bar{H}=2$.}\label{fig:utrap_4x1_dynamics}
\end{figure}

We now examine the bifurcations of equilibria that occur in flat-bottomed trapezoidal cross-sections with respect to changes in the curvature parameter $\epsilon$ for a small particle $\alpha=0.05$, paying specific attention to how these change as $\delta$ increases.
The methodology is analogous to that in section~\ref{sec:strap_bifur} and comparisons with results obtained for the family of symmetric trapezoidal cross-sections having the same aspect ratio ($W/\bar{H}=4$) in figure~\ref{fig:strap_4x1_dynamics} will be particularly useful.
As before, we choose to plot with respect to $\kappa$ rather than $\epsilon$ (noting $\kappa=32000\epsilon$ given fixed $\alpha=0.05$) so that the results can be more readily interpreted for similar particle sizes.

Figure~\ref{fig:utrap_4x1_dynamics} shows the bifurcations which occur within several flat-bottomed trapezoidal cross-sections.
The top row, corresponding to a rectangular cross-section ($\delta=0$), is provided for reference (and is identical to the top row of figure~\ref{fig:strap_4x1_dynamics}).
The second row of figure~\ref{fig:utrap_4x1_dynamics} shows the bifurcations that occur for shape parameter $\delta=0.1$.
The left plot, showing the horizontal coordinates of equilibria vs $\kappa$, is barely distinguishable from the symmetric trapezoidal case (see figure~\ref{fig:strap_4x1_dynamics}, second row), there being just a slight difference in the $r$-coordinate of the two equilibria which make up the stable pair in a neighbourhood of $\kappa=10$ and again near $\kappa=100$.
In contrast, the middle plot, showing the vertical coordinates of equilibria vs $\kappa$, illustrates some subtle differences compared to the bifurcations observed in the vertically symmetric trapezoidal cross-sections. 
However, these differences relate to the saddle and stable equilibria near the side walls which ultimately have little impact on the dynamics of most particles.

Upon increasing the shape parameter to $\delta=0.2$ (middle row of figure~\ref{fig:utrap_4x1_dynamics}) we begin to observe several changes in comparison to the equilibria observed in the equivalent symmetric trapezoidal cross-section.
The first observation is the absence of one stable equilibrium and one saddle equilibrium near the inside wall (which existed for values of $\kappa\lesssim10$ for smaller $\delta$, and persisted in the vertically symmetric trapezoidal cross-section for $\delta=0.2$ but not $\delta=0.3$).
The second observation is the increasing offset in the $r$-coordinate of the two equilibria in the stable pair over the range of $\kappa$.
This helps to distinguish the two distinct folds that occur near to $\kappa=100$, and thus distinguish the two cusp bifurcations that occur with respect to the parameter $\delta$.
Of course, there are also two such folds in the vertically symmetric trapezoidal cross-sections, but those were only distinguishable by their $z$-coordinates.
Examining the $z$-coordinates of equilibria further highlights the effect of the increasing asymmetry of the cross-section, and also the absence of the two aforementioned equilibria.
We'll refrain from discussing these in detail as they ultimately don't affect the migration dynamics of most particles.

With further increases to $\delta$ (bottom two rows of figure~\ref{fig:utrap_4x1_dynamics}) the changes described above become further exaggerated. 
For instance, there is an increasing offset in the $r$-coordinate of the two equilibria in the stable pair, the range of $\kappa$ covered by the fold increases, and there is increasing asymmetry in the $z$-coordinates of equilibria.
Observe that the stable equilibrium that occurs in the upper portion of the cross-section must undergo larger changes in $z$ with respect to $\kappa$ as $\delta$ increases due to the increasing slope of the nearby top wall.
One thing that remains consistent from $\delta=0.2$ to $\delta=0.4$ is that there are no fundamentally new equilibria or bifurcations that occur (which is qualitatively consistent with the vertically symmetric trapezoidal cases for $\delta=0.3,0.4$).

\subsection{Comparison of dynamics for several particle sizes}

\begin{figure}
\centering
\includegraphics{{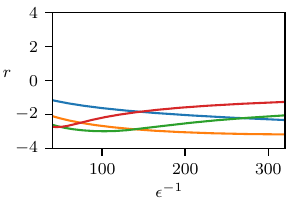}} \,
\includegraphics{{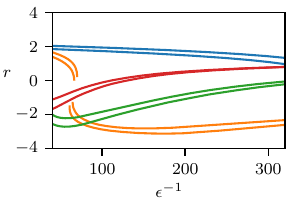}} \,
\includegraphics{{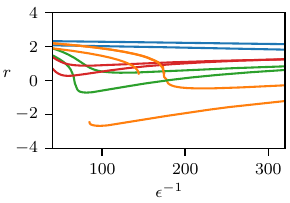}} \\
\includegraphics{{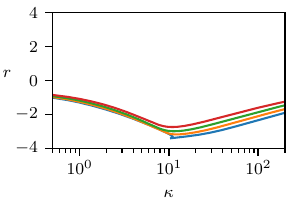}} \,
\includegraphics{{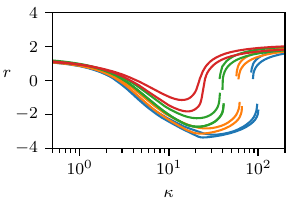}} \,
\includegraphics{{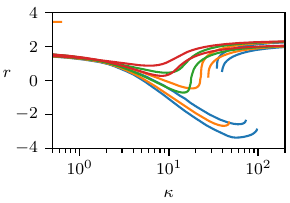}} \\
\caption{Change in horizontal location of the stable equilibrium pair with respect to $\epsilon^{-1}$ (top row) and $\kappa$ (bottom row) for flat-bottomed trapezoidal cross-sections with shape parameters $\delta=0,0.2,0.4$ (left to right) and aspect ratio $W/\bar{H}=4$. 
The colour of each curve corresponds to particle sizes $\alpha=0.05$ (blue), $\alpha=0.1$ (orange), $\alpha=0.15$ (green), $\alpha=0.2$ (red).}\label{fig:utrap_trends}
\end{figure}

Figure~\ref{fig:utrap_trends} shows the subtle changes in the horizontal location of the two equilibria making up the stable pair due to differences in particle size $\alpha$.
The top row shows the $r$-coordinates vs $\epsilon^{-1}$ while the bottom row shows the $r$-coordinates vs $\kappa$. 
The columns, left to right, correspond to flat-bottomed trapezoidal cross-sections with $\delta=0,0.2,0.4$, respectively, and $W/\bar{H}=4$.
We will discuss these in comparison with the analogous results for the vertically symmetric trapezoidal cross-sections (having the same $W/\bar{H}$ and $\delta$) of figures~\ref{fig:strap_eps_trends} and \ref{fig:strap_kap_trends}.

The $\delta=0$ case is provided for reference and shows how focusing locations are confined to a relatively small portion of the duct width in the case of rectangular cross-sections.
For $\delta=0.2$ the results are remarkably similar to the results obtained for the analogous symmetric trapezoidal cross-section, apart from the slight offset in horizontal locations of the two equilibria in each pair due to the absence of vertical symmetry.
Consequently, most of the observations made in relation to the vertically symmetric cross-section continue to apply here. 
In particular, over a large range of practical $\epsilon^{-1}$ values we observe that all four particle sizes can be quite well separated.
Moreover, on examining the trends with respect to $\kappa$, the three smaller particle sizes appear to have just undergone a cusp bifurcation (with respect to $\delta$).

Moving on to the case of $\delta=0.4$ the larger offset in horizontal location of the two equilibria making up the stable pair is clear but there remain some broader qualitative similarities to the results obtained for the equivalent symmetric trapezoidal cross-section.
Observe from the the plot of horizontal location vs $\epsilon$ (top right), most equilibria are in the region $r\gtrsim 0$, i.e. there is  significantly less separation between the different $\alpha$ compared to the case $\delta=0.2$.
Additionally, the offset in horizontal location of the two equilibria for the $\alpha=0.1$ (orange) particle is particularly pronounced over this range of $\epsilon^{-1}$.
One interesting feature is that for the $\alpha=0.1$ (orange) particle, the lower stable equilibrium features a fold in its curve but the upper stable equilibrium is a continuous curve.
This indicates there has been a second cusp bifurcation that re-joins the curve corresponding to the upper equilibrium for the $\alpha=0.1$ particle (i.e. between $\delta=0.2$ and $\delta=0.4$).
Looking at the plot of the horizontal location vs $\kappa$ we see that, despite the offsets due to broken symmetry, the curves still come together reasonably well for both sufficiently small and sufficiently large values of $\kappa$.

\section{Conclusions}\label{sec:conclusions}

In this work we have applied our model of inertial particle migration in curved microfluidic devices to undertake a thorough dynamical study of particle migration in two exemplary families of trapezoidal cross-sections.
While the underlying model has a number of limitations in order to make it practical to implement, e.g. by assuming that the Dean number and particle Reynolds number are small and neglecting any particle/fluid acceleration effects, this study provides a number of practical insights into the use of trapezoidal cross-sections in microfluidic device design.
We have paid particular attention to observations that may be important in the context of size-based particle separation, since this is one area where curved ducts with trapezoidal cross-sections have been actively explored in the experimental literature.

While there is much that could be summarised in a broad study such as this, there are two particular findings that we highlight here.
The first is that the most significant change in the bifurcations associated with the change in trapezoidal shape parameter appears to occur near to $\delta=0.2$.
Specifically, we observe cusp bifurcations which present as a fold that develops in the curves describing the horizontal location of stable equilibria vs the $\kappa$ parameter (recalling $\kappa=4\epsilon/\alpha^3$ where $\epsilon=\bar{H}/(2R)$ is the non-dimensionalised curvature of the duct centreline and $\alpha=2a/\bar{H}$ is the non-dimensionalised particle size).
By exploiting the steepening of these curves near the onset of the fold, devices may potentially be designed to achieve a large separation of particles having only a small relative size difference.
Additionally, we illustrated that cross-sections with $\delta=0.2$ provide very good separation of all four particle sizes considered in this study over a relatively large range of achievable bend radii $R=\epsilon^{-1}\bar{H}/2$.
For larger values of $\delta$ the stable equilibria begin to concentrate nearer to the outer wall, implying smaller degrees of separation between particles of different size.

The second finding is that there is much qualitative overlap in the dynamics of the stable pairs observed in the vertically symmetric trapezoidal cross-sections and flat-bottomed trapezoidal cross-sections.
While the latter exhibits an offset in the horizontal location of the two equilibrium in the stable pair, this offset is generally small for $\delta\leq 0.2$ and the qualitative behaviour is otherwise quite well approximated by the vertically symmetric case.
This suggests that, while flat-bottomed trapezoids may be more practical to produce and use in experiments, we can obtain a reasonably good understanding by studying their vertically symmetric counterparts, which affords some computational efficiencies.

There are several directions in which this work may be extended.
Of course, there are ample opportunities to explore additional aspect ratios and even entirely different families of cross-sectional shapes.
The accurate estimation of the fields $\hat{L}_\ast,\hat{S}_\ast,\hat{D}_\ast$ takes quite a bit of computation and so it is currently not feasible to exhaustively explore every imaginable cross-sectional shape, but there may be other interesting one-parameter families to explore.
Recent work has produced an efficient way to estimate the stable equilibria of $\hat{L}_\ast$ in straight duct geometries \cite{ChristensenEtal2022}, and this might be combined with an estimate of $\hat{S}_\ast$ to provide an alternative approach for investigating a greater variety of cross-sections.
There are also opportunities to lift some of the restrictions of the current model as has been done in other studies involving rectangular cross-sections, e.g. the consideration of non-neutrally buoyant particles \cite{HardingStokes2020}, or an extension to moderate Dean numbers \cite{HardingStokes2023}.
Ultimately, combining these ideas within a tool to facilitate optimal device design for targeted applications would be of great benefit to end-users of microfluidics research.


\section*{Acknowledgments}
Many of the results were computed using resources provided by the R\=apoi HPC at Victoria University of Wellington. 
We thank Juan Patino Echeverria for his exploratory work examining a toy model of inertial migration in curved trapezoidal ducts as part of a summer research project at Victoria University of Wellington over the summer of 2021--2022.

\appendix

\section{Additional results for a large $\delta$ and a negative $\delta$}\label{app:extraD}

Here we briefly provide some results for a cross-section with $W/\bar{H}=2,4$ and $\delta=-0.1,0.9$.
The inclusion of $\delta=-0.1$ illustrates that stable focusing equilibria become more concentrated near the inside wall for $\delta<0$, while the inclusion of $\delta=0.9$ illustrates what happens when the cross-section is almost triangular in shape.

\begin{figure}
\hspace{2cm}\includegraphics{{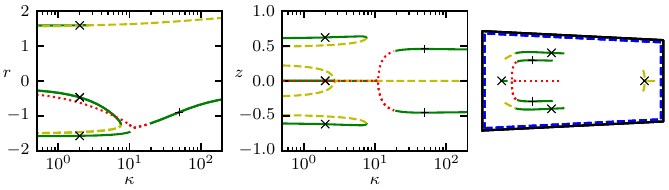}} \par
\hspace{2cm}\includegraphics{{figures_pdf/strap_kappa_dynamics_2x1_a05_D00_v2-eps-converted-to.pdf}} \par
\hspace{2cm}\includegraphics{{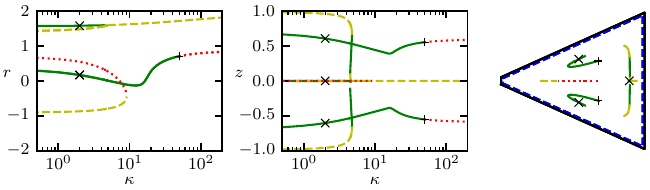}}
\caption{
Dynamics associated with the fixed points of \eqref{eqn:ode_model} for particle size $\alpha=0.05$, aspect ratio $W/\bar{H}=2$, and trapezoidal shape parameter $\delta=-0.1,0,0.9$ (top to bottom). 
Line styles denote stability: green solid for stable, yellow dashed for saddle, red dotted for unstable.
The left and centre column show the horizontal and vertical coordinate, respectively, of fixed points vs $\kappa$ (or equivalently $32000\epsilon$)
The right column shows the path followed by fixed points within the cross-section as $\kappa$ changes. 
The cross and plus markers illustrate the location of stable fixed points at the specific values of $\kappa=2,50$, respectively.
Cross-sectional coordinates have been re-scaled so that $\bar{H}=2$.
}\label{fig:strap_2x1_extraD}
\end{figure}

Figure \ref{fig:strap_2x1_extraD} shows the bifurcations that occur for the particle size $\alpha=0.05$ in the case $W/\bar{H}=2$, analogous to figure \ref{fig:strap_2x1_dynamics}.
The case $\delta=0$ (middle row) is included again as a point of reference.
For $\delta=-0.1$ (top row) observe that the stable equilibrium pair is generally located closer to the inside wall of the duct.
For $\delta=0.9$ (bottom row), observe that the horizontal position of the stable equilibrium pair does not move as much as $\kappa$ changes.
Another interesting feature is the very brief existence of a second stable equilibria pair near the outside wall around $\kappa\approx 4$ for $\delta=0.9$.
Additionally, observe the existence of regions where there are no stable equilibria, such as $\kappa\gtrsim50$ for $\delta=0.9$ and $10\lesssim\kappa\lesssim20$ for $\delta=-0.1$, which ultimately means there are stable limit cycles in these parameter ranges.

\begin{figure}
\hspace{2cm}\includegraphics{{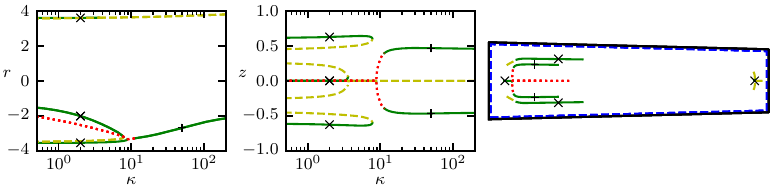}}

\hspace{2cm}\includegraphics{{figures_pdf/strap_kappa_dynamics_4x1_a05_D00_v2-eps-converted-to.pdf}}

\hspace{2cm}\includegraphics{{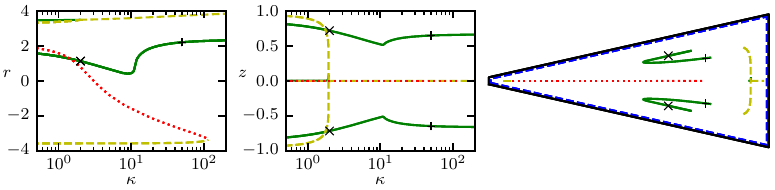}}
\caption{
Dynamics associated with the fixed points of \eqref{eqn:ode_model} for particle size $\alpha=0.05$, aspect ratio $W/\bar{H}=4$, and trapezoidal shape parameter $\delta=-0.1,0,0.9$ (top to bottom). 
Line styles denote stability: green solid for stable, yellow dashed for saddle, red dotted for unstable.
The left and centre column show the horizontal and vertical coordinate, respectively, of fixed points vs $\kappa$ (or equivalently $32000\epsilon$)
The right column shows the path followed by fixed points within the cross-section as $\kappa$ changes. 
The cross and plus markers illustrate the location of stable fixed points at the specific values of $\kappa=2,50$, respectively.
Cross-sectional coordinates have been re-scaled so that $\bar{H}=2$.
}\label{fig:strap_4x1_extraD}
\end{figure}

Figure \ref{fig:strap_4x1_extraD} shows the bifurcations that occur for the particle size $\alpha=0.05$ in the case $W/\bar{H}=4$, analogous to figure \ref{fig:strap_4x1_dynamics}.
For $\delta=-0.1$ (top row) observe that the stable equilibrium pair is generally located closer to the inside wall of the duct.
This supports the intuition that trapezoidal ducts which are taller at the inside wall are not conducive to particle separation.
It is noteworthy that there is a small region around $\kappa\approx10$ for which there are no stable equilibria, which ultimately means there is a stable limit cycle.
For $\delta=0.9$ (bottom row), observe that the stable equilibrium pair traverses a smaller range in the horizontal direction as $\kappa$ varies.
Additionally, as this equilibrium pair follows a single unbroken curve, there must be a second cusp bifurcation between $\delta=0.4$ and $\delta=0.9$ which ``straightens out'' the fold produced by the cusp bifurcation between $\delta=0.1$ and $\delta=0.2$.


\bibliographystyle{siamplain}
\bibliography{ms}

\end{document}